\documentclass[12pt]{article}
 \usepackage{amsmath}
 \usepackage{amsthm}
 \usepackage{amsfonts}
 \usepackage{graphicx,psfrag,epsf}
 \usepackage{enumerate}
 \usepackage{times}
 \usepackage{caption}
 \usepackage{subcaption}
 \usepackage{natbib}
 \usepackage{multirow}
 \usepackage{bm,algorithm} 
 \usepackage{url} 
 \usepackage{comment}
 \newcommand{\blind}{0}
 \usepackage{setspace}

 \usepackage[pagewise]{lineno}
 
 \def\spacingset#1{\renewcommand{\baselinestretch}%
 {#1}\small\normalsize} \spacingset{1}

 \usepackage{geometry}
 \usepackage{tikz}
 \usetikzlibrary{arrows,shapes.arrows,shapes.geometric,shapes.multipart,backgrounds,decorations.pathmorphing,positioning,fit,automata}

 \usepackage{xr}

 \makeatletter
 \newcommand*{\addFileDependency}[1]{
   \typeout{(#1)}
   \@addtofilelist{#1}
   \IfFileExists{#1}{}{\typeout{No file #1.}}
 }
 \makeatother
 
 \newcommand*{\myexternaldocument}[1]{%
     \externaldocument{#1}%
     \addFileDependency{#1.tex}%
     \addFileDependency{#1.aux}%
 }

 \myexternaldocument{supplement}

 \usepackage{epsf,pstricks,pst-node}
 \usepackage{float}
 \usepackage[outdir=./]{epstopdf}
 \usepackage{verbatim} 

 \newcommand{\bea}{\begin{eqnarray*}}
 \newcommand{\eea}{\end{eqnarray*}}
 \newcommand{\be}{\begin{eqnarray}}
 \newcommand{\ee}{\end{eqnarray}}
 \newcommand{\bay}{\begin{array}}
 \newcommand{\eay}{\end{array}}
 \newcommand{\bi}{\begin{itemize}}
 \newcommand{\ei}{\end{itemize}}
 \newcommand{\ben}{\begin{enumerate}}
 \newcommand{\een}{\end{enumerate}}
 \newcommand{\bcen}{\begin{center}}
 \newcommand{\ecen}{\end{center}}

 \usepackage[symbol]{footmisc}

 \DeclareOldFontCommand{\bf}{\normalfont\bfseries}{\mathbf}

 \allowdisplaybreaks

 \date{} 
 
\begin{document}
  \if0\blind
  {
   \title{\bf    Statistical Learning Methods for Neuroimaging Data Analysis with Applications }

  \maketitle
 \begin{center}
   \author{\large Hongtu Zhu$^{1}$,  Tengfei Li$^2$, and Bingxin Zhao$^3$ \\
 \vspace{10pt}
   $^1$Departments of Biostatistics, Statistics, Genetics, and Computer Science and Biomedical Research Imaging Center,  University of North Carolina, Chapel Hill \\
  $^2$Departments of Radiology and Biomedical Research Imaging Center,  University of North Carolina, Chapel Hill \\
  $^3$Department of Statistics and Data Science, University of Pennsylvania
  }
 \end{center}
 \newpage
 } \fi

  \if1\blind
  {
   \title{\bf Statistical Learning Methods for Neuroimaging Data Analysis with Applications}
   \maketitle
 } \fi

 \vspace{-50pt}
 \spacingset{1.7} 

 \begin{abstract}
 The aim of this paper is to provide a comprehensive review of statistical challenges in neuroimaging data analysis from neuroimaging techniques to large-scale neuroimaging studies to statistical learning methods.  
 We briefly review eight popular neuroimaging techniques and their potential applications in neuroscience research and clinical translation. 
 We delineate the four common themes of  neuroimaging data and review major image processing analysis methods for 
 processing neuroimaging data at the individual level.  We briefly review four large-scale neuroimaging-related studies and a
 consortium on imaging genomics and discuss four common themes of neuroimaging data analysis at the population level.  
 We review nine  major population-based statistical analysis methods  and  their associated statistical challenges and present recent progress in statistical methodology to address these challenges.
 \end{abstract} 

 \begin{quote}
 \textbf{Keywords}: causal pathway,    heterogeneity,    image processing analysis,  
 neuroimaging techniques, population-based statistical analysis, study design. 
 \end{quote}

 \vspace{-30pt}

\section{Introduction}
\label{Introduction}

Neuroimaging refers to the process of producing images of  
 the structure, function, or pharmacology of the central nervous system (CNS). 
It has been a dynamic and evolving field   with (A1) the development  of   new acquisition techniques, (A2) the collection of various  neuroimaging data in clinical settings and  medical research, and (A3)
 the development of statistical learning (SL) methods.  
For (A1), popular neuroimaging techniques include structural magnetic resonance imaging (sMRI),  
functional magnetic resonance imaging (fMRI), diffusion weighted imaging (DWI),
computerized tomography (CT), 
positron emission tomography (PET), 
electroencephalography (EEG), magnetoencephalography (MEG), and 
functional near-infrared spectroscopy (fNIRS).
These techniques were developed to  measure   specific tracers in  CNS, that are directly and indirectly associated with brain structure and function. 
For instance, PET  delineates how an injected radioactive tracer (e.g., Fluorodeoxyglucose (FDG))
moves and accumulates in the brain, whereas fMRI 
measures an indirect tracer, called
 the concentration of deoxyhaemoglobin, in the flow downstream of the activated neurons caused by 
 brain’s activity.  
   The developments of SL methods for individual  neuroimaging data raise serious challenges for existing statistical methods   due to  four  common   themes consisting of   
   \textbf{(CT1)} \textit{complex brain objects}, \textbf{(CT2)} \textit{complex spatio-temporal structures},  \textbf{(CT3)} \textit{extremely  high dimensionality}, and 
        \textbf{(CT4)} \textit{heterogeneity  within subjects and across groups}.

\begin{figure}[ht]
\centering
\includegraphics[height=3in,width=5 in]{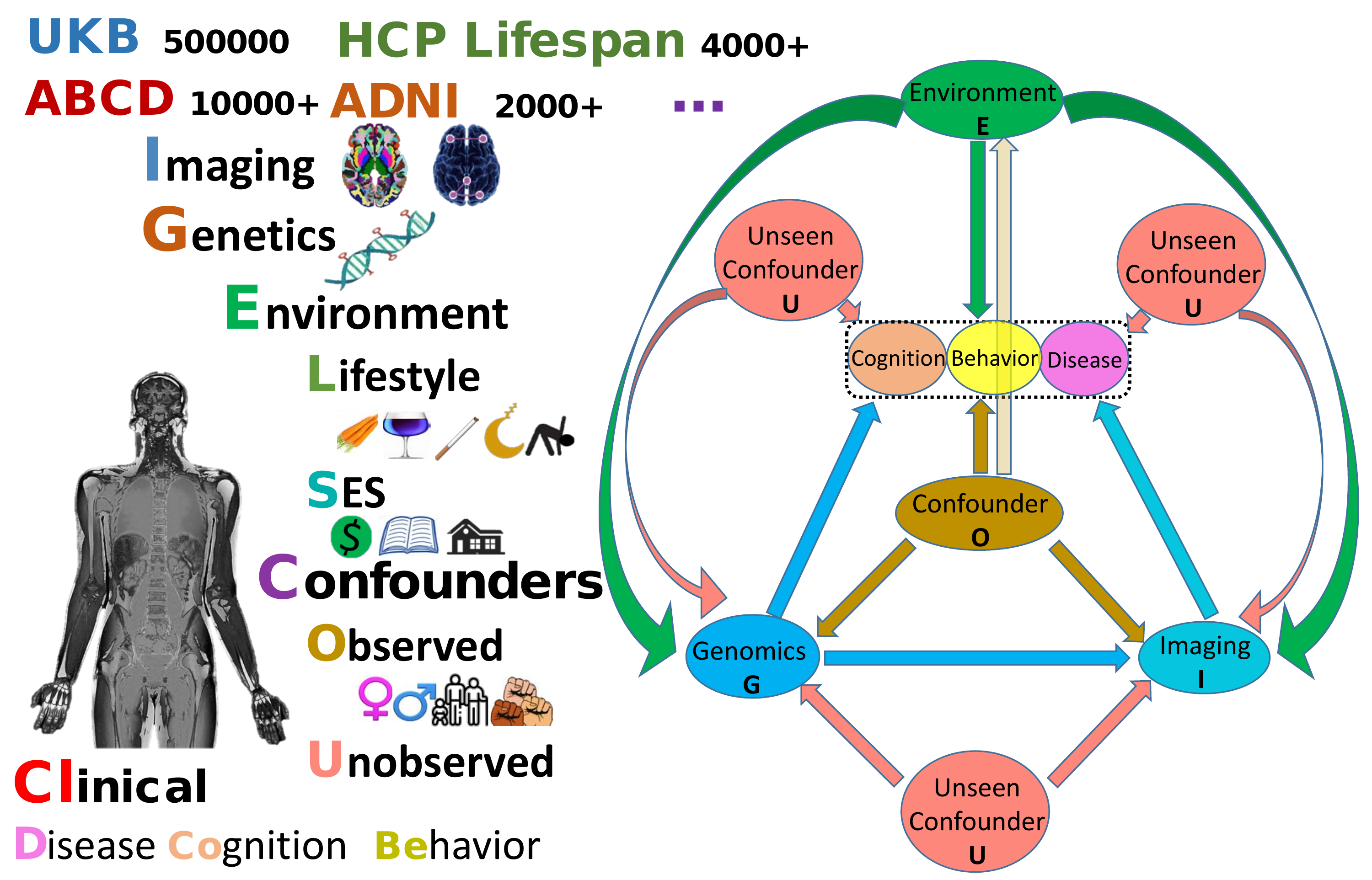}
\caption{Left panel: Major data types from different domains   in several representative large-scale biomedical studies; Right panel: A dynamic  causal model  for delineating causal genetic-imaging-clinical (CGIC) pathway confounded with environmental factors and unobserved confounders.}\label{fig:integration}
\end{figure}

 For (A2),  in recent years, huge amounts of neuroimaging data have been collected  in health care,   biomedical research studies, and clinical trials.  First, neuroimaging  
 has the potential to improve clinical care   for  diagnosis and prognosis in various brain-related diseases, such as     dementia, sleep disorders, and schizophrenia.  
 Some typical uses of neuroimaging include  
identifying the effects of brain-related  diseases (e.g., stroke or glioblastoma), 
locating cysts and tumors, and 
finding swelling and bleeding, among others.  
Second,  
many large-scale  biomedical studies  have collected/are collecting  massive neuroimaging data (e.g., sMRI, DWI, and fMRI) with high spatial and/or temporal resolution as well as other
complex information (e.g., genomics and health factors) in order to map the human brain connectome  for 
 understanding the pathophysiology of brain-related disorders, the progress of neuropsychiatric and neurodegenerative disorders,  the
normal brain development, and the diagnosis of brain cancer, among others. In the last two decades, there are at least 
three  pioneering neuroimaging-related studies, including    Alzheimer's Disease
Neuroimaging Initiative (ADNI)~ (http://www.adni-info.org/) \citep{weiner2010alzheimer},
 the Human Connectome Project (HCP)~  (http://humanconnectome.org/consortia/) \citep{van2013wu},
and   the {UK Biobank (UKB) study}~ (https://www.ukbiobank.ac.uk/) \citep{miller2016multimodal}. They represent major advances and innovations in    acquisition  protocols, analysis pipelines,    data management,  experimental design, and  sample size. 
The left panel of Figure \ref{fig:integration} shows  multi-view data across different domains (e.g., imaging, genetics, or environmental factors) in some large-scale biomedical studies. Third,   
neuroimaging biomarkers have many uses 
 in clinical trials for  drug development  in  neurological and  psychiatric  disorders \citep{schwarz2021use}.  
These uses include a screening tool for 
selecting   trial participants,  
a tool to establish biodistribution, 
target engagement and pharmacodynamic activity, 
a means for monitoring safety, and
an evidence measure of disease modification.
   The developments of SL methods for clinical translation and  large-scale neuroimaging-related  studies raise serious challenges to existing statistical methods   due to the four additional themes of  \textbf{(CT5)} \textit{sampling bias},     
   \textbf{(CT6)}  \textit{complex missing patterns},      
     \textbf{(CT7)} \textit{complex data objects},
     and  
   \textbf{(CT8)} \textit{complicated causal pathways in brain disorders}. 
 
 For (A3),  there is a large literature on the development of  SL methods  for neuroimaging data analysis (NDA) in order to correlate multi-type data from different domains across multiple studies, eventually establishing a dynamic causal pathway (e.g., the causal genetic-imaging-clinical (CGIC) pathway in the right panel of Figure \ref{fig:integration}) linking genetics to brain (or neuroimaging) phenotypes to clinical outcomes confounded with health factors.   
 These SL methods can be   categorized  into two categories including image  processing analysis (IPA)  at individual level and population-based statistical  analysis (PSA) for a sample of subjects.   
 We further group various   IPA methods into deconvolution and structure learning \citep{sotiras2013deformable,li2019computational,zhou2021review,shen2017deep,park2003super,yi2019generative}. Deconvolution methods primarily include image reconstruction and image enhancement.   Structure learning methods mainly include image segmentation and image registration. Due to (CT1)-(CT4)  and the lack of high-quality annotation datasets, it is very challenging  to  develop   `good' IPA pipelines  to extract a relatively small number of  image phenotypes (IPs) with high  repeatability and reproducibility for  both individual health care and  PSA.  
We also group various   PSA methods into nine main categories, including  study design, statistical parametric mapping,  object oriented   data analysis,   dimensional reduction methods,  data integration,  imputation methods,  
  predictive models, imaging genetics, and   causal discovery   
 \citep{ombao2016handbook,Nathoo2019,shen2019brain,smith2018statistical,nichols2017best,rathore2017review}. 
Due to (CT1)-(CT8), each category has its own statistical challenges, requiring specific statistical methodology to address them.  
 However,   the development of scalable PSA methods     has fallen seriously behind the technological advances in neuroimaging techniques, causing difficulty in translating research findings to clinical practice.

 \section{A Review of Neuroimaging Techniques and Uses}  
   
\begin{figure}[ht]
\centering
\includegraphics[height=5 in,width=5in]{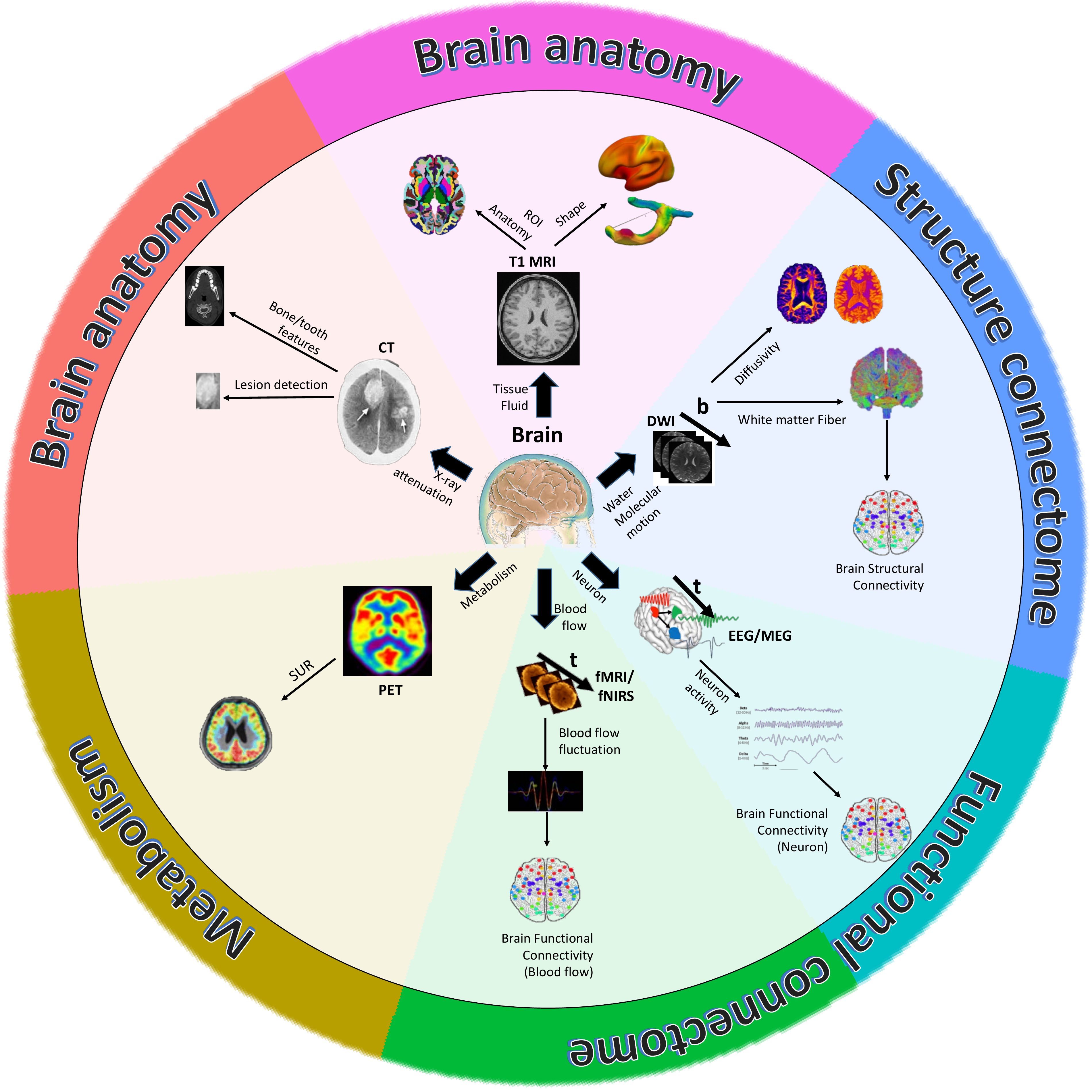}
\caption{Roles of different imaging modalities for extracting various types of features.}\label{fig: modality}
\end{figure}

  We briefly review  eight neuroimaging techniques below and summarize them in supplementary Table 1.  For each image modality, we describe its tracer, data dimension, features,  main uses,  and several key softwares \citep{smith2010introduction}.
  Figure \ref{fig: modality} also presents different neuroimaging modalities and the different types of features they extract.  
\begin{itemize}   
\item{} {\it Structural magnetic resonance imaging (sMRI)}  measures the  fluid characteristics of different tissues (gray and white matter), creating high-resolution (0.5mm-1mm)  images with a strong gray/white matter contrast and many anatomical details.
It allows us to 
qualitatively and quantitatively measure the development and change of  cortical and subcortical structures in terms of  both size and shape  in the brain. 
Some  sMRI derived measurements include cortical thickness, cortical  folding,
 sulcal depth, voxel-based morphometry,  and 
regional volumes and shape.  
sMRI has been  widely used  for diagnosis, staging, and follow-up of disease in  clinics and brain development in research.  

\item{} {\it Diffusion weighted MRI (DWI)}  
 measures the  Brownian motion of water molecules within voxels, creating  images with a relatively low spatial resolution of  1.25-3 mm and multiple $b-$values and tens to a few hundreds of diffusion directions that can reveal  microscopic details about tissue architecture and map white matter trajectories in the brain.
 It allows us to  qualitatively and quantitatively measure white matter (WM)  trajectories and water diffusivity along these trajectories {\it in vivo}. 
Some  DWI derived measurements include invariant measures (e.g., fractional anisotropy) along WM trajectories (or in WM regions)  and 
their related 
weighted and binary network metrics (e.g., the counts of streamlines connecting all WM region pairs).  
DWI has been  used  for delineating tumors, suspected acute ischemic brain injury,  intracranial infections, masses, trauma, and edema in clinics and mapping structural connectome in research.

\item{} {\it Functional MRI (fMRI)} primarily measures the blood-oxygen-level-dependent (BOLD)  responses in blood flow  associated with brain function, creating  images with a typical  spatial resolution of  3-4 mm, a typical temporal resolution of 1-3 s, and  hundreds of time points
that 
 can map metabolic function and  neuronal activity.
  It allows us to  indirectly and non-invasively  measure brain functions  under  specific tasks, resting state, and naturalistic paradigms. Thus, fMRI consists of task based fMRI (tfMRI) and resting state fMRI (rsfMRI).  
  Some  fMRI derived measurements include voxel-wise activation patterns (e.g., beta images), region-based activation and interaction patterns, and
weighted and binary network metrics (e.g., the correlation matrix for  all region-of-interest pairs).    
  fMRI  has been used for 
  brain activity mapping under different tasks, 
 brain abnormalities detection, and 
 pre-operative  mapping of brain functions. 
 \item{} {\it Positron emission tomography (PET)} 
  measures  
  emissions from  radioactive tracers (e.g., 18F-FDG), creating images with 
a spatial resolution of   4-5 mm, and a poor temporal resolution 
(tens of seconds to several minutes)  that can reveal 
  tissue's metabolism (e.g., flow, oxygen, and glucose metabolism). 
  It allows us to  qualitatively and quantitatively measure   the physiology  and anatomy  of brain as well as its biochemical properties. 
{  Some  PET-derived measurements include voxel-wise activation patterns (e.g., standard uptake ratio (SUR) images) and region-based activation and interaction patterns. }  
    PET has been used for mapping brain functions and  detecting abnormalities in brain neurophysiology and neurochemistry associated with Alzheimer's Disease,  anxiety, and stroke. 
 
 \item{} {\it  Computerized tomography (CT)} 
   measures   X-ray attenuations by different tissues inside the body, creating images with a high spatial resolution of tens of nanometers    that can non-destructively  reveal internal details (e.g.,  soft tissues or  bones) of  organs.   
  It allows us to  qualitatively and quantitatively measure   
  brain tissue and brain structures, such as skull and blood vessels.  
  However, 
  CT as a radiation diagnostic technique can cause adverse effects, including harmful tissue reactions and cancer. 
{  Some  CT derived measurements include  local and regional volumetric and thickness measures. }
    CT has been used for  diagnosing a range of conditions, such as 
abnormal blood vessels, brain atrophy, hemorrhage, swelling, stroke, and tumors. 
\item{} {\it Electroencephalography (EEG)} measures the electrical field  produced by neuron electrical activity, creating 
an electrogram of the electrical activity on the scalp with a high temporal resolution of millisecond or less. 
It allows us to indirectly and non-invasively measure synchronous dendritic activity of cortical pyramidal neurons. 
However, EEG has a poor spatial resolution of 5-40 cm$^3$, which depends on the number of electrodes ranging from tens to a few hundreds. Some EEG derived measures include   
 event-related potentials (e.g., stimulus onset) linked to   an event, connectivity measures, network measures,  and  the type of neural oscillations   in the spectral content of EEG, including   
  delta (1–4 Hz), theta (4–8 Hz), alpha activity (8–12 Hz), beta (13–30 Hz), low gamma (30–70 Hz),  and high gamma (70–150 Hz). 
  EEG has been used  for diagnosing and treating 
brain tumors, brain damage, brain dysfunction, 
sleep disorders, anxiety, epilepsy, 
inflammation, and 
stroke. 
\item{} {\it Magnetoencephalography (MEG)} measures the magnetic field  produced by neuron electrical activity, creating 
an electrogram of the electrical activity  with a high temporal resolution of millisecond or less and a moderate spatial resolution of a few millimeters.   Compared with scalp EEG, MEG uses very sensitive magnetometers to   indirectly and non-invasively measuring the tangential components of  post-synaptic intracellular currents in the dendrites of neurons.  MEG and EEG share similar derived measures. 
MEG has been used  for identifying the functional areas of the brain, including centers of sensory, motor, language, and memory activities, and for mapping 
the precise location of the source of epileptic seizures. 

\item{}  {\it Functional near-infrared spectroscopy (fNIRS)} 
uses infrared light (650–900 nm) to measure changes in cortical BOLD response associated with brain function, creating an electrogram of BOLD signals with a high temporal resolution of milliseconds and a spatial resolution of millimeters below cortical surface.  fNIRS shares 
  similar derived measures with EEG and fMRI.  
fNIRS has been used for  
studying normal and pathological brain physiology and investigating behavioral and cognitive development in infants and children.
 \end{itemize}

There is a great interest in developing different integration methods to fuse multimodel neuroimaging together  \citep{tulay2019multimodal}, since
no single modality is able to completely delineate the
complex dynamics of brain physiology and pathology. 
It allows us 
to borrow complementary information from different modalities, 
leading  a  comprehensive picture
of the brain under different clinical conditions, different tasks,
resting state, and normal development. 
Three categories of multimodel neuroimaging include  structural–structural combinations, 
  functional–functional combinations, and  structural–functional combinations.  For instance, 
the fMRI/EEG integration as a functional-functional combination  improves both spatial and temporal resolution, while 
cross-validating findings across different scales. 
A simultaneous CT-MRI scanner
  as a special case of structural-structural combinations is to 
 integrate   high contrast resolution of MRI with high spatial resolution of
CT. 
Structural-functional combinations, including 
EEG/sMRI, PET/CT, and PET/MRI,  
link anatomical structure
with functional dynamics, 
improving mapping the
anatomical basis of brain functions   and simulating brain dynamics. 
 Furthermore, scientists proposed whole-brain models by combining anatomical networks extracted from  DWI/sMRI with local dynamics extracted from fMRI/EEG/MEG  and metabolism extracted from PET  \citep{deco2015rethinking}. 
Those  whole-brain models usually consist of  three basic elements, including brain parcellation (e.g., the   HCP-MMP in  \cite{glasser2016multi}), 
anatomical connectivity matrix for the human connectome,
and local dynamics for the activity of each brain region and interaction terms with other regions.  
 \section{Image Processing Analysis (IPA) Methods} 
  
  We   discuss the four common themes of neuroimaging data,  
  review  existing major IPA methods for processing neuroimaging data, 
  and  delineate major statistical 
  challenges associated with IPA.  
  
  \subsection{Common Themes (CT1)-(CT4)} 
  
We discuss  four  common  themes of  neuroimaging data as follows.

   \textbf{(CT1) Complex Brain Objects.} All neuroimaging modalities, including those in Section 2, are developed to  indirectly  (or directly) measure the structure and function of the cerebrum, cerebellum, brain stem, diencephalon (thalamus and hypothalamus),  limbic system, 
reticular activating system, and ventricular system in the human brain.  
For instance, the cerebrum   is part of the forebrain, consisting of the cerebral cortex  of gray matter  as the outer layer and  white matter in  the inner layer. It is responsible for processing 
language, motor function, memory, vision,  personality, and other cognitive functions. 
The cerebral cortex  consists of frontal lobe,   temporal lobe, parietal lobe, and  occipital lobe, 
while its surface  is made up of gyri and sulci.
Moreover, the human brain uses neurons as information messengers to send 
electrical impulses and chemical signals to different brain regions and body in order to    
control biological functions and react to environmental changes.  
Moreover,  there are two sets of blood vessels, including  the vertebral arteries and the carotid arteries, that supply blood and oxygen to the brain. These objects in the brain are the targets of different neuroimaging modalities. 

     \textbf{(CT2) Complex Spatiotemporal Structures.} There are three different  spatiotemporal structures, including spatial and temporal resolutions,  spatio-temporal smoothness, and spatiotemporal correlation. 
     In Section 2, we have discussed different spatial and temporal resolution ranges for the eight neuroimaging techniques.    
     In general, higher  spatial (or temporal) resolution leads to better spatial (or temporal) localization, but in some cases (e.g., DWI), higher spatial resolution  decreases signal-to-noise ratio. 
     Due to the intrinsic smooth structure of different brain regions discussed in (CT1), neuroimaging data 
    is expected to contain spatially contiguous regions or
effect regions with relatively sharp edges, showing 
 locally  strong spatio-temporal smoothness and  spatiotemporal correlation.  Moreover,   
     long-range temporal  correlations among different brain regions may be caused by  respiration, cardiac rhythm, and cognitive processes. 
     
     \textbf{(CT3) Extremely  High Dimensionality.} Both raw  neuroimaging data and  extracted feature data can be extremely high dimensional even for a single subject. For instance,  for a single subject,   the number of 3-dimensional (3D) DWI images  varies from several tens to a few hundreds, and the extracted feature data includes 3-dimensional images of
 various diffusion-related quantities (e.g., diffusion tensors and fractional anisotropy), a
whole-brain tractographic data set (which  can contain more than 1,000,000 streamlines), 
diffusion properties along white matter bundles, and structural connectivity network metrics. 
For a single subject,   the number of 3D  tfMRIs  is about several hundreds and the extracted feature data includes 3D activation patterns, region-based activation and interaction patterns, and
weighted and binary network metrics.

\begin{figure}[ht]
\centering
\includegraphics[height=3in,width=5 in]{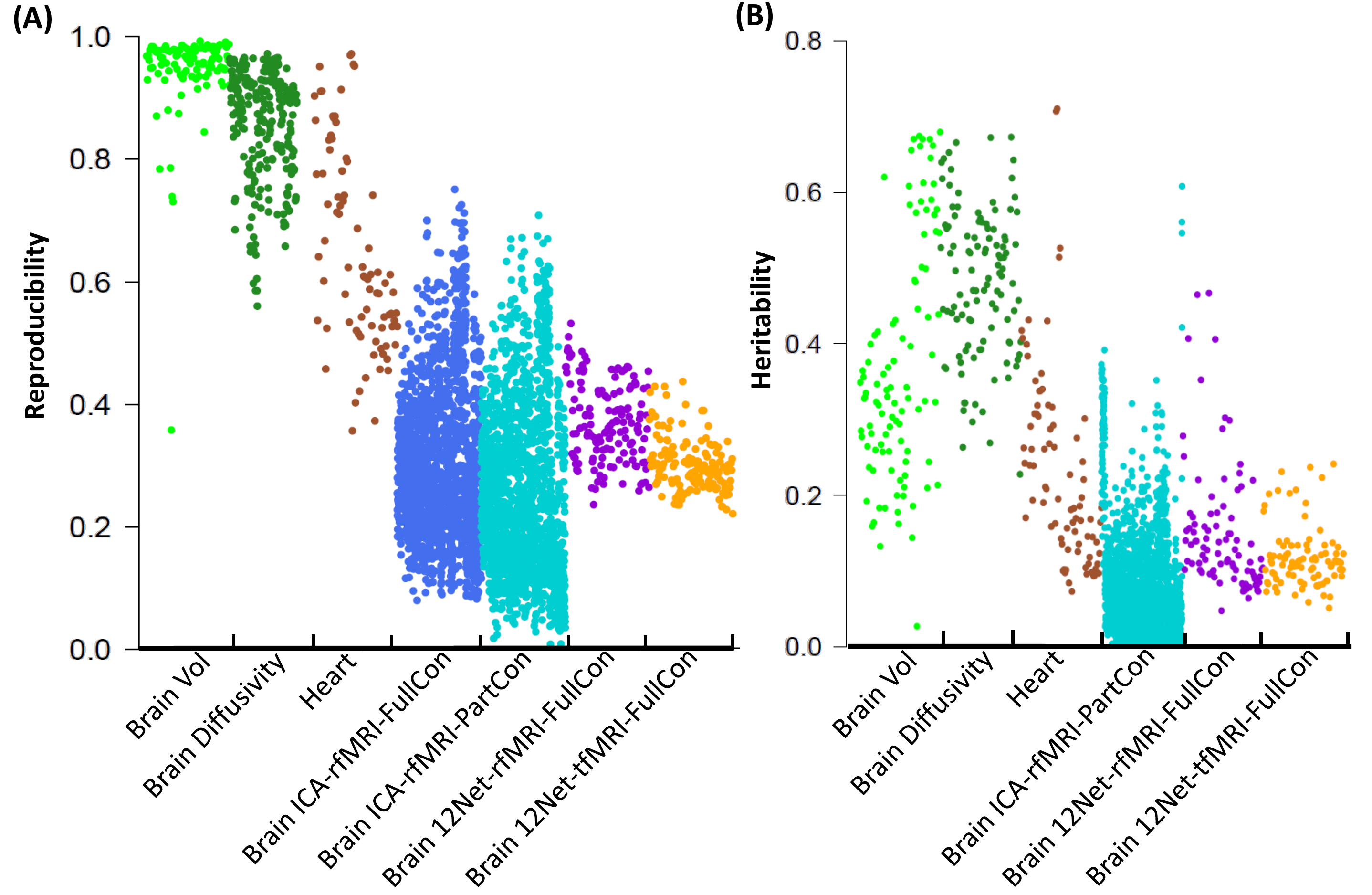}
\caption{The reproducibility (left panel) and heritability (right panel) of seven  different categories of imaging traits based on UKB including brain regional volume, brain diffusivity parameters, heart MRI traits, brain independent component analysis (ICA)-based resting-state fMRI full and partial connectivity, 12 network-based brain rsfMRI full connectivity, and 12 network-based brain tfMRI full connectivity.}\label{fig: icc}
\end{figure}

      \textbf{(CT4)
     Heterogeneity within Individual Subjects and across Centers/Studies}   A neuroimaging data may be written as \begin{equation} \label{eq1}
     I=f(\mbox{brain} (\mbox{age}, \mbox{gene}, \mbox{race}, \mbox{disease}, \mbox{other factors}), \mbox{device},  \mbox{acquisition parameters},  \mbox{noises}), 
     \end{equation} 
     where noises contain all kinds of noise components (e.g., thermal noise or motion) \citep{smith2018statistical} and brain includes both brain structural and functional components.  
     Model (\ref{eq1}) emphasizes two important facts that (i)  neuroimage data represent a mixture of different components introduced by  
     brain,   device, acquisition parameters, and different noises and (ii) brain changes may be caused by age, genes, race, disease,  and  other factors (e.g.,  stimulus,   life style,  or environmental factors).  
     The effect of device, acquisition parameters, and noises in $I$ 
       can be  larger than the effect of brain changes caused by predictors of interest. 
     For a single subject in a short time window,  it is expected that structural images are much more stable than  functional images even in the same scanner, whereas one may observe visible  differences in the same type of structural  images  acquired in two different scanners. 
     A sensible neuroimaging modality requires that brain changes caused by a specific condition are large relative to the variability caused by noises, acquisition parameters, and device. 
     { Figure \ref{fig: icc}A presents the reproducibility using intraclass correlation coefficient (ICC) values of  imaging phenotypes based on the UKB test-retest dataset. We observe that the brain  and heart structural traits have much larger reproducibility than the brain functional traits, suggesting the complexity and variability of  brain function.  }

    Any novel IPA methods  for   neuroimaging data need to account for some/all of the four themes (CT1)-(CT4) discussed above.  We review two categories of  IPA methods  including deconvolution and structural learning in the existing literature below.

  \subsection{IPA: Deconvolution}
  
  We use ``deconvolution" to represent all computational and statistical methods for reconstructing image data  of interest from recorded imaging  signals with various noise components.  We can further  categorize all deconvolution methods into  the image reconstruction and enhancement processes \citep{park2003super,hansen2015image}. 
  
The {\it  image reconstruction process} 
for neuroimage data aims to  reconstruct clinically interpretable images from  raw  data acquired by neuroimaging devices. 
For instance, MRI data are acquired in k-space and a 
specific  image reconstruction process is needed to generate MRI images in  image space. Several key methods for  MRI reconstruction  
include 
noise pre-whitening, zero filling in k-space, raw data filtering, Fourier transforms, and phased array coil combination \citep{hansen2015image}. 
Recently,    compressed sensing algorithms and deep learning (DL) methods play a critical role in fast MRI acquisition and reconstruction   
\citep{chen2022ai,lustig2008compressed}.  
Furthermore, most neuroimage data in the image space still need additional reconstruction  in order to estimate local features of interest 
in the human brain.   
Some examples include   diffusion tensors for DWI, cortical surface for sMRI,  white matter fiber bundles for DWI, and hemodynamic response functions for fMRI and fNIS  
\citep{ombao2016handbook,zhu2007statistical,yeh2021mapping,seghouane2019robust,liu2008reconstruction,jenkinson2012fsl}.

The {\it image enhancement  process} 
for neuroimage data is to improve the quality of the generated images for better presentation and analysis.  
Popular enhancement
tasks include denoising, super-resolution,  bias
field correction, and   harmonization.  Among them, 
 bias
field correction  and   harmonization were proposed to correct for two major confounders including devices and artifacts in noises described  in Model  (\ref{eq1}).  
Specifically,  
  bias field in image data  refers to the presence of a low frequency intensity nonuniformity, representing  a potential confounder
  in various image analysis tasks, such as tissue segmentation  \citep{song2017review}.  
 Various bias correction methods (e.g., nonparametric nonuniform intensity normalization (N3) algorithm)  fall into  prospective and retrospective approaches     according to different sources of bias field and features used in bias correction \citep{song2017review}. 
 Harmonization in imaging data aims to correct significant inter- and intra-site variability even within individual subjects, which may be caused by   hardware, reconstruction process, and acquisition parameters.   Such   variability are much more profound across different subjects in multi-site and multi-study neuroimaging datasets.   Therefore, there is a great interest in the development of various harmonization methods, including surrogate variable approach, meta analysis, mega analysis, removal of artificial voxel effect by linear regression,  phantome-based harmonization, DL,   or combined association test (Combat),  for correcting inter- and intra-site variability in neuroimaging datasets \citep{yu2018statistical,chen2021privacy}. 
 See Section \ref{DIDI} for further details.

  \subsection{IPA: Structure Learning}

 We use ``structure learning" to include all computational and statistical methods for extracting  signals of interest from reconstructed  imaging data.  We can further  categorize  structural learning methods into  the image segmentation and registration processes \citep{sotiras2013deformable,li2019computational,zhou2021review,shen2017deep,bharati2022deep,miller2001group,Grenander2007,hesamian2019deep,srivastava2016functional}.

 The {\it image segmentation process} 
for neuroimage data aims to label reconstructed neuroimaging data into  meaningful subgroups for clinical and scientific tasks, including the quantification of brain development, the localization of pathology, surgical planning, and image-guided interventions. 
Existing image segmentation methods can be roughly clustered into traditional segmentation techniques (e.g., intensity-based methods or surface-based methods), machine learning approaches,  and deep-learning ones, such as fully connected networks (FCNs) and U-nets \citep{isensee2021nnu,hesamian2019deep}. 
  Major   neuroimage segmentation  tasks include  skull stripping, cortical and subcortical structures segmentation, white matter tract parcellation, functional parcellation, and   lesion localization
    \citep{kalavathi2016methods,eickhoff2018imaging,fischl2012freesurfer,wasserthal2018tractseg,havaei2017brain,isensee2021nnu}.  
   Performing these tasks allows us to extract a wealth of important features, including  local properties of brain structures;
    short-, median-, and long-range structural and functional connectivity patterns; and  
      structural and functional markers, while 
  addressing (CT1)-(CT4),

Segmentation tasks have at least three important applications. 
First, they greatly compress the  dimensionality of neuroimaging data as detailed in (CT3), while providing strong 
biological interpretation.
Second,  refined     
brain structural and functional parcellations  greatly improve our understanding of the organizational principles behind
the human brain across multiple regions, multiple scales, and multiple tasks.  
Third,     an important  clinical application of image segmentation is computer-aided detection and diagnosis  for  
  localizing   lesions  and then classifying them into a specific lesion type \citep{zhou2021review}.

  The {\it image registration process} for neuroimage data aims to 
 transform 
the spatial coordinates of  neuroimage data within individual subjects  and/or across different subjects into the same 
  coordinate system of an atlas \citep{bharati2022deep,miller2001group,Grenander2007,hesamian2019deep,srivastava2016functional}. 
 Some important  applications of registration include the construction of brain atlas, 
multimodal fusion, the quantification of brain development, population analysis, longitudinal analysis,  automated image segmentation, shape
analysis, and the localization of pathology. 
 Most image registration algorithms have three major components including (i)  the similarity measure, (ii) the transformation
model, and (iii) the optimization process. 
The similarity measures can be either intensity-based  (e.g., mutual information or correlation metrics) 
or feature-based (e.g., distances between image features such as points, lines, and contours).
The  transformation models can be categorized into    rigid (translations and rotations), affine, homographies, and deformations models.
Deformation models  \citep{sotiras2013deformable} can be further grouped into   physics-based models based on   a physical model, interpolation-based methods, and knowledge-based approaches, leading to ill-posed problems. It requires 
imposing implicit and explicit regularization constraints,  such as hard constraints, 
 topology preservation,  volume preservation, and  
rigidity constraints. 
Recently, we witness a growing interest in the development of  DL-based image registration methods, such as 
deep iterative registration, deep supervised registration, and deep unsupervised registration \citep{bharati2022deep}. 
These DL-based models hold great promise for using  a single forward
calculation to complete   registration 
  within few seconds,    
while showing comparable accuracy with conventional methods.

As an example,  we consider 
the construction of imaging-based human brain atlases as   
one of the most important applications of registration.  
Cartographic 
approaches have been widely used  to 
create  anatomical atlases (e.g., Brodmann’s 
map and Dejerine’s map) based 
on post-mortem tissues,  establishing   
spatial correspondences between a coordinate and a brain structure.
Recently, we witness a tremendous evolution  of   human brain atlases (e.g., Yeo-Network, AHB,  or HCP-MMP)  \citep{glasser2016multi,eickhoff2018imaging,toga2006towards,nowinski2021evolution}  due to   
the availability of  many advanced imaging techniques, brain mapping methods, large-scale neuroimaging datasets, and 
registration methods, among others. Various criteria have been applied for human brain atlases, including brain architecture, 
functional activities, anatomical and functional connectivity, abnormality, genetic and protein information,  cell type,  lifespan, spatio-temporal scales, ethnicity, and multiple modalities,  
among others.  In the near future, 
modern human brain atlases may provide  
an integrative and comprehensive description of brain structure and function  in  large populations, across different scales, age,  gender, behavioral tasks, ethical groups,  disease states, and imaging modalities. 

\subsection{A Generic Statistical Model for IPA}

 We discuss a generic statistical model for IPA, including denoising, super-resolution,  reconstruction, segmentation, and registration. 
 First, we consider image reconstruction. Suppose that we observe $\{ ({\bf x}_i, I_i): i=1, \ldots, n\}$, where  $I_i$ and ${\bf x}_i$ are, respectively,  an imaging vector and 
 a predictor vector, which may  depend on the imaging device, acquisition parameters, and observable confounders in noise components.  It is assumed that 
 $I_i$ given ${\bf x}_i$ follows a probability distribution $p(I_i|h({\bf x}_i, \theta), \sigma)$, where $\theta$  is  a vector of parameters (or functions), $\sigma$ is a vector of nuisance parameters,  and $h(\cdot, \cdot)$ is a vector of  functions.  
 Let's consider two examples. First, we consider the raw sMRI data in k-space. In this case, $I_i$ is the complex magnetic resonance imaging (MRI) measurement in k-space, ${\bf x}_i$ 
 includes its $(k_x, k_y)$ coordinate and other MRI scanner parameters, $n$ is the total number of observations in k-space, and $\theta$ is the sMRI in image space. Second, we consider the DWI data. In this case, $I_i$ is   DWI, ${\bf x}_i$ includes $b$-values and diffusion directions, $n$ is the total number of DWI, and $\theta$ is the image of diffusion tensors.

The primary interest of many deconvolution methods is to  estimate $\theta$  by maximizing  
\begin{equation} \label{eq2}
     L_n(\theta)=\sum_{i=1}^n\log p(I_i| h({\bf x}_i, \theta), \sigma)+R_1(h({\bf x}_i, \theta))+R_2(\theta, \sigma),  
     \end{equation} 
     where $R_1(\cdot)$ and $R_2(\cdot)$ are two regularization terms based on  prior information, such as sparsity and spatio-temporal structures in (CT1) and (CT2).  
     As an illustration, we discuss how to construct $\log p(I_i| h({\bf x}_i, \theta), \sigma)$ in equation (\ref{eq2})   for image denoising by using weighted loss functions. Many denoising methods  solve 
  a weighted loss function by incorporating signals in 
  the neighboring locations of each location.   A further refinement is to  build a sequence of  increasing neighborhood sizes  and then sequentially fit the weighted loss function in equation (\ref{eq2}) to estimate $\theta$ as size starts  from the smallest size to the largest size, while borrowing information from the previous sizes   \citep{Polzehl:Spokoiny:2000,Zhu:2011}. In this case, 
  $L_n(\theta)$ may implicitly depends on all observations in  the neighboring locations of each location, so it is strongly dependent on both location and neighborhood size.  
  Specifically,    we estimate $\theta$ in equation  (\ref{eq2}) at the smallest size, denoted as $\hat\theta_{(0)}$,   and then use adaptive smoothing methods to sequentially calculate $\hat\theta_{(s_k)}$ for $s_0=0<s_1<\ldots<s_K$, while preserving  spatial smoothness and edges 
   \citep{Buades:Coll:Morel:2005}.

Both image segmentation and registration can be also formulated as special cases of equation  (\ref{eq2}). For image segmentation, 
 ${\bf x}_i$ and ${ I}_i$ are, respectively, input image data for segmentation and output segmentation results, $n$ is the number of annotated image data, and $R_1(\cdot)$ may be a spatio-temporal regularization term. 
 For image registration, we consider registering a pair of images with 
 ${\bf x}_i$ and ${I}_i$ being source image and target image, respectively. 
In this case, $n=1$, $h({\bf x}_i, \theta)={\bf x}_i(T_i(s))$ with $T_i(\cdot)$ being a transformation model, 
$\log p(I_i| h({\bf x}_i, \theta), \sigma)$ is a matching criterion chosen to match $(I_i, {\bf x}_i)$,  
and $R_1({\bf x}_i(T_i(s)))$ is imposed on $T_i(\cdot)$ to induce certain constraints (e.g., diffeomorphism) \citep{bharati2022deep,miller2001group,Grenander2007,hesamian2019deep,srivastava2016functional}.

  \subsection{Challenges} 
  
  We have briefly reviewed four major IPA techniques including reconstruction, enhancement, segmentation, and registration,  which are the key building blocks of most neuroimage preprocessing pipelines, but each of them  requires substantial efforts on validation, which can be a daunting and difficult task. For instance,  
most neuroimage segmentation methods suffer from a major data bottleneck (or barrier) for validation,  
 even though    DL-based methods have significantly improved segmentation accuracy over traditional methods. 
 Specifically,  there are no single, publicly available, high-quality neuroimaging   datasets with detailed annotation information that cover a large spectrum of segmentation tasks  in neuroimaging research, greatly limiting the translation of segmentation methods to the clinic. 
 In contrast, publicly available datasets and environments (e.g., the ImageNet) played a vital role in    the development of DL methods for computer vision  problems and the successes of ‘narrow AI’ systems, such as  DeepMind’s AlphaGo. 
 Several methodological attempts to partially address the data bottleneck for validation include unsupervised learning,  self-supervised learning, 
weakly supervised learning, data augmentation, patch-wise training,  and  transfer learning \citep{zhou2021review,isensee2021nnu,hesamian2019deep}. However, there is a great need to accomplish several key developments in order to address the data bottleneck including  
   the development of 
good annotation protocols for  major segmentation tasks,  the collection of high-quality datasets covering a wide range of settings as discussed in (CT4),  the use of  
active learning and reinforcement learning  \citep{budd2021survey,zhou2021deep}, and  a comprehensive evaluation system for image segmentation and registration.  Similar comments are also valid for validating most image registration methods.

As an illustration, we consider a comprehensive DWI preprocessing pipeline consisting of 
(i) fiber orientation reconstruction, (ii) WM tracking, (iii) WM parcellation, (iv)  WM registration, (v) extraction of diffusion properties along WM and structural connectivity metrics, (vi) visualization, and (vii) statistical analysis.    
Although  major technical advancements have been made in Steps (i)-(vii) in the last decade,     
     steps (ii)-(iv) still face  major technical barriers.   
 Specifically,  multiple tractography challenges reveal that 
       most state-of-the-art algorithms produce   many more false  than valid WM bundles \citep{schilling2019limits,schilling2022prevalence}, leading to erroneous  structural connectivity metrics.  
   Those false WM bundles are mainly caused by the limitation of DWI and the complexity of WM structure as discussed in (CT1). 
       Moreover,   a recent open call for segmenting 
  14 white matter fascicles based on the same sets of streamlines obtained from six  subjects \citep{schilling2021tractography} reveals that 
there is a large variability across  57 different state-of-the-art segmentation protocols and techniques for such call. 
Such variability is mainly caused by  the complexity of WM structure as discussed in (CT1) and the lack of good validation data sets 
in addition to the limitations of existing clustering techniques.  
The variability  in WM tracking and parcellation  greatly affects    downstream WM structural connectivity metrics extraction and quantification \citep{schilling2021tractography}. 
Another technical barrier is that 
existing WM
registration algorithms not only suffer from pinching effects for transforming WM bundles to the WM bundle atlas \citep{srivastava2016functional}, but also  
 largely ignore the
diffusion property information along fiber tracts \citep{zhang2018mapping}, causing a local misalignment issue
among those diffusion property functions.  
In contrast,  the method of tract-based
spatial statistics (TBSS) \citep{Smith:2006}, which projects WM diffusion properties onto a whole brain WM skeleton, is a robust approach with high reproducibility (Figure \ref{fig: icc}),  but TBSS 
 does not have individual fiber tract 
specificity.




  \section{ Large-scale Neuroimaging-related Studies}  
  
\begin{figure}[ht]
\centering
\includegraphics[height=5in,width=5 in]{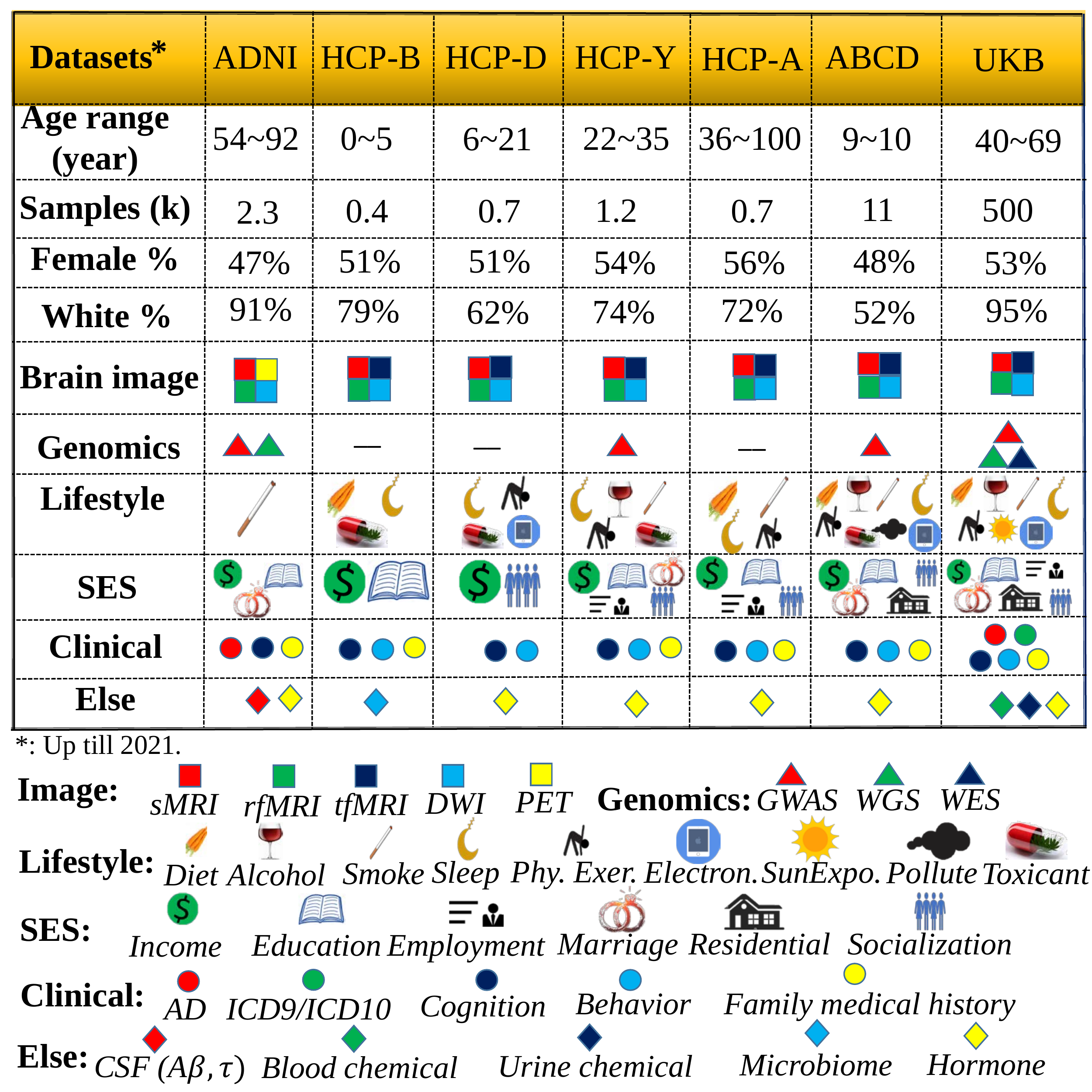}
\caption{Some summary information for the ADNI, HCP, ABCD and UKB studies.}\label{fig: datatable}
\end{figure}

  We  witness the exponential increase in the collection of neuroimaging data in many large-scale biomedical studies (e.g., UKB) 
   in the last decade primarily due to huge investment from  different funding agencies and private sectors \citep{miller2016multimodal,littlejohns2020uk}. The number of subjects in a neuroimaging study increases from several tens  in most neuroimaging-related studies thirty years ago  to more than 10,000 in several studies lately.   
   Besides neuroimaging data, those
    large-scale biomedical   studies  have collected/are collecting   other data types, including genetic data,  behavioral data, environmental factors,   and clinical outcomes in order to better understand  the progress of  neuropsychiatric  disorders, neurological disorders, and stroke, and    normal brain development,   among many others. Recently, several large 
consortiums  have been  formed to enhance collaborations on neuroimaging and imaging genomics among researchers across the world.  
  We brief review four large-scale neuroimaging-related studies, whose detailed information is included in 
    Figure \ref{fig: datatable}, and a consortium on imaging genomics, called ENIGMA.  
  
  
\subsection{ADNI}

 The overall goal of ADNI is to validate potentially useful biomarkers for AD clinical treatment trials \citep{weiner2017alzheimer}. 
 ADNI is a multisite, prospective clinical study and   
actively supports the investigation and development of treatments that may slow or stop the progression of Alzheimer's disease (AD) \footnote{ADNI, \url{https://adni.loni.usc.edu/study-design/}}.
Researchers across 63 sites in the US and Canada have been tracking the progression of AD through clinical, imaging, genetic and biospecimen biomarkers, starting from normal aging, early mild cognitive impairment (EMCI), late mild cognitive impairment (LMCI) to dementia or AD. The recruitment for ADNI was designed to mimic a clinical trial population, where the participants were generally well educated, mostly white, and with a high proportion of 
{\textit {APOE4}} 
carriers among the MCI and AD groups, which is consistent with subjects in MCI and AD clinical trials. The ADNI participants do not represent “typical” subjects across the old population since  the proportion of AD cases (23\%) is much higher than the prevalence of AD in the US. 
Up till 2022, the ADNI study has collected 2,723 participants aged above 40 in four phases including  ADNI1, ADNI2, ADNIGO, and ADNI3.  Genomic biosample and genotyping were collected for all subjects, enabling polygenic risk scores and gene pathway- and
network-based metrics for prediction of disease progression. Cerebrospinal fluid (CSF) biomarkers, including the $A\beta$, t-tau and p-tau, were provided. For imaging data, ADNI1 focused primarily on sMRI and PET to study brain changes in brain morphology and metabolism with AD.  ADNIGO and ADNI2  added a high-resolution coronal T2, perfusion MRI, DWI, and resting-state fMRI (rsfMRI).  ADNI3 includes seven
sequences (sMRI, FLAIR, T2*GRE, DWI, rsfMRI, ASL perfusion MRI, and a high resolution coronal T2 fast spin echo) in all subjects. ADNI3 aims to  study how tau PET, CSF biomarker, and functional imaging affect treatment, promote the development of new immunoassay
platforms and mass spectroscopy techniques to improve
the reliability of CSF analysis, and deepen our understanding of the progression and pathophysiology of AD \citep{weiner2017alzheimer}. 
As a multimodal longitudinal AD-targeted study,  ADNI has provided a myriad of important insights on various aspects of AD, emphasizing AD pathophysiology and and disease progression.

\subsection{HCP}
The original Human Connectome Project (HCP-Y) gathered data from 1,200 healthy young adults aged 21-35, including young adult sibships of average size 3–4, to build a high-quality data set that can be comparable with other populations. The primary goals of HCP  include (i)  building a ``network map'' that will shed light on the anatomical and functional connectivity within the healthy human brain, (ii) promoting the understanding of inter-individual variability of brain circuits to behavior,  (iii) facilitating research into brain disorders, such as  autism, AD, and schizophrenia, and (iv) making all data freely available to the scientific community \citep{van2013wu, bookheimer2019lifespan}. Now,  it has been extended to a number of studies on healthy humans ranging from birth to nonagenarians and beyond \footnote{HCP lifespan, \url{https://www.humanconnectome.org/lifespan-studies/}}, aiming at  mapping  neural systems to underlying cognition and behavior across the life span. Those studies will include  HCP-B (HCP babies: age 0-5; 500 subjects), HCP-D (HCP development: age 5-21; 1,350 subjects), and HCP-A (HCP aging: age 36-100+; 1,200+ subjects). All HCP studies are hybrid cross-sectional and longitudinal cohorts, which recruited participants according to specific inclusion and exclusion criteria, such as the age range, birth weight,  no major diagnosed diseases, and informed consent, with longitudinal follow-up observations for subsets of participants. Such recruitment method ensures that the samples reflect the racial/ethnic and socioeconomic diversity of the US Census. 
The HCP collected various imaging modalities including DWI, rsfMRI, tfMRI, T1- and T2-weighted sMRI, and   MEG/EEG. Domains of cognition, emotion, motor function, and sensation were also collected, while different major factors relevant to brain development, aging, cognition and behavior were collected for different age phases. For example,  HCP-A collected the vascular burden (e.g., obesity, hypertension, smoking), risk gene status (e.g., {\textit {APOE}}), hormonal status, and lifestyle factors (e.g., depression, sleep patterns,  social/community engagement, and adversity) \citep{bookheimer2019lifespan}. For HCP-Y participants, genotyping data are available across 2 million SNPs from 1142 study participants, while for HCP study participants outside of the HCP-Y cohort, samples will be assayed on several SNP regions of interest. Follow-up samples will also be collected for longitudinal assessment. Two major advantages of HCP  include maximized resolution of imaging data and overall data quality for multi-modal imaging. 

 \subsection{ABCD}

The ABCD study is the largest prospective longitudinal study of brain development and child health in the United States, which has recruited approximately 11,880 children aged 9-10 years old from 21 research sites  and is following them for 10 years into early adulthood \footnote{ABCD, \url{https://abcdstudy.org/about/}}. Its initial goal was to examine risk and resiliency factors associated with the development of substance use, and then expanded far beyond, into identifying the underlying biospecimens, neural alterations, and environmental factors,  and their contributions to the development of behavior, brain function, and other mental and physical outcomes throughout adolescence \citep{karcher2021abcd}. The ABCD adopted multi-stage probability sampling strategy to recruit eligible children to reflect as best as possible, the sociodemographic variation of the US population. However, more neuroimaging research centers were located in urban areas, leading to a potential under-representation of rural youth. 
The ABCD study  covers  personal information, family structure, family socioeconomic status,  medical history,  mental/behavioral performance, lifestyle (physical activity, sleep, diet), substance use (both self-reported and screening: alcohol, nicotine, cannabis, caffeine, cocaine, and marijuana), exposure (air pollution and lead),   neuroimaging data (sMRI, DWI, and rsfMRI and tfMRI), and genotyping data. At baseline and year 1 follow-up sessions, biological breath, saliva, urine, and hair samples were collected from youth and genotyping were performed from saliva and blood DNA sample for 11,601 participants.  ABCD provides a comprehensive platform for investigating gene-environmental effects on children brain development. 

\subsection{UKB}
UKB is a very large prospective cohort study that have recruited over 500,000 individuals aged 40 and 69 from 22 centers across the United Kingdom. 
It aims to inspire the imaginations of health researchers around the world to meet the challenge of greater understanding, prevention, and treatment of a range of serious illnesses \footnote{UK Biobank, \url{https://www.ukbiobank.ac.uk/}}. Extensive phenotypic and genotypic details about its participants were collected, including data from questionnaires focused on health and
lifestyle, physical measures, sample assays, accelerometry, multimodal imaging, genome-wide genotyping and longitudinal follow-up for a wide range of health-related outcomes. The UKB imaging study is by far the largest  multi-modal imaging study in the world, with over 50,000 participants having undergone assessments \citep{littlejohns2020uk}, including brain sMRI, brain fMRI, brain DWI, body MRI, low-dose X-ray bone and joint scans, and ultrasound of the carotid arteries. The genotype data, whole exome sequencing, and whole genome sequencing data for 500,000, 470,000, and 200,000 participants are  available to researchers up till 2022, respectively. Finally, over 19,155 diagnostic terms has been collected including hospitalization episode statistics (HES) and recorded using the International Classification of Diseases, Tenth Revision (ICD-10) codes. There is expected to be 20 years of longitudinal follow-up on the participants, and the identification of disease risk factors should increase over time with emerging clinical outcomes.  The UKB data set provides a unique opportunity for  uncovering the genetic bases of brain   structure and function,    aging,  and various diseases. 
For the recruitment procedure, postal invitations were sent to 9,238,453 individuals aged 40–69 years old, who lived within 25 miles from one of 22 assessment centres in the UK. With a response rate of 5.5\%, there is significant evidence of selection biases, including the higher socio-economic status, better education and health of the UKB sample compared to the general population, leading to debates on the generalizability of UKB findings. Nevertheless, as reported in \cite{batty2019generalisability}, many findings from UKB appear to be generalizable to England and Scotland.

  \subsection{ENIGMA}
  The Enhancing NeuroImaging Genetics through Meta-Analysis (ENIGMA) Consortium is a global alliance of over 1,400 scientists across 43 countries in the fields of imaging genomics, neurology, and psychiatry, studying  a range of large-scale  human brain studies that integrate data based on sMRI, DWI, fMRI, genetic data and many patient populations from over 70 institutions worldwide \citep{thompson2020enigma}. Launched in December 2009, the initial goal of the ENIGMA was to discover the impact of genetic factors on brain systems by integrating the two big data sources---neuroimaging and genetics.  The major goals of ENIGMA include\footnote{ENIGMA, \url{https://enigma.ini.usc.edu/}} (i) pushing forward the field of imaging genetics, (ii) ensuring promising and reproducible findings, (iii) sharing data, ideas, methods, algorithms and other information, and (iv) training new investigators. The consortium consists of over 50 working groups (WGs), including diagnosis-based, normal variation-based, and method-based WGs.  In 2014, ENIMGA considered nine targeted disorders: schizophrenia, bipolar disorder, major depressive disorder, obsessive-compulsive disorder, attention-deficit/hyperactivity disorder, autism spectrum disorders, substance use disorders, 22q11.2 deletion syndrome, and the effects of the human immunodeficiency virus on the brain. Following this, additional work groups focusing on specific disorders were established, including anxiety disorders, suicidal thoughts and behavior, sleep and insomnia, eating disorders, irritability, antisocial behavior, and dissociative identity disorder. Besides the diagnosis-based WGs, normal-variation WGs study the brain lifespan development, normal aging, gender difference, sleep patterns, and early onset psychosis, whereas  method-based WGs span over developing innovative pipelines on producing DWI measures, anatomical shape measures, and data harmonization. Up till now,  ENIGMA has stood out for its great impact in promoting robustness and reproducibility, setting methodological standards, and driving new discoveries in neuroscience research and clinical translation.

  \section{Population-based Statistical Analysis (PSA) Methods} 
  
  We discuss  four common themes of NDA in  those large-scale biomedical studies as discussed in Section 4, review existing major PSA methods for   NDA,  and discuss major statistical challenges associated with PSA.  

\subsection{Common Themes (CT5)-(CT8)}

Although we have discussed the common themes (CT1)-(CT4) of neuroimaging data, four more themes as detailed below arise from the joint analysis of big neuroimage data and other related variables 
from many large-scale biomedical studies, such as UKB and ENIGMA.  

{\bf (CT5) Sampling Bias.} The most important issue in   NDA is how to appropriately address potential sampling bias introduced at  design and data collection stages. Some common types of sampling bias include 
undercoverage, observer bias, voluntary response bias, survivorship bias,  recall bias, and exclusion bias \citep{riffenburgh2012statistics}. 
A direct consequence of sampling bias is that 
the sample in a study is not a representative sample of a target population.   
 Sampling bias can have profound effects on downstream data analysis as well as on the generalizability and fairness  (e.g., sex, race, or age) of conclusions drawn from statistical  models. Although the issue of sampling bias is prevalent in neuroimaging research,  
 it has been largely ignored in the  medical imaging literature until recently  \citep{roberts2021common,batty2019generalisability}. Understanding how to appropriately 
 deal with sampling bias requires development of specific strategies 
 in the design and collection stages as well as statistical models to explicitly model the sample selection process \citep{thompson2012sampling}. 

{\bf  (CT6) Complex Missing Patterns.} 
 Missing data  frequently encountered in large-scale neuroimaging studies are caused by  various reasons, including  missing by design, faulty scanning, attrition in longitudinal studies, mis-entry,  and non-responses in surveys, among others. For a single variable with missing data, 
these are three types of  missingness, including missing completely at random (MCAR), missing at random (MAR), and missing not at random (MNAR). 
  Simply ignoring missing observations and improperly imputing them may lead to efficiency loss and introduce spurious correlations. 
Additional challenges also arise in handling missing data in large-scale neuroimaging related studies. 
For instance, variables with different missing patterns often occur in the same neuroimaging  study, while high-dimensional image data are block-wise missing  either within individual studies or across different studies. 
Little progress has been ever made on how to  appropriately  integrate information across different domains from multiple heterogeneous studies in the presence of block-wise missing data \citep{xiang2014bi}, even though  
 there is a large literature on handling missing entries of low-dimensional clinical outcomes \citep{LittleRubin2002,ibrahim2009missing}.

{\bf (CT7) Complex Data Objects.} 
Complex data objects in curved spaces frequently arise in the process of extracting meaningful features with strong biological  interpretation from neuroimaging data. 
Some examples of data objects include  
planar shapes, symmetric 
positive definite matrices,  matrix Lie groups, tree-structured data,    the Grassmann manifolds,  deformation fields,  connectivity graphs, functional connectivity graphs, diffusivity properties along WM bundles,  and the shape representations of cortical and subcortical 
structures, among others. 
Most of these complex data objects   are 
inherently nonlinear as well as high-dimensional (or even infinite-dimensional), so many 
traditional statistical techniques, including semiparametric and nonparametric regression, growth curve models, clustering, classification,  correlation, and dimension reduction,  are often not be directly 
applicable  to them \citep{Dryden1998,marron2021object,huckemann2021data,cornea2017regression,srivastava2016functional,wang2016functional,dubey2020functional}. 
The undertaking of efficient analysis of complex data objects as well as variables obtained from other domains
presents major statistical and computational challenges.

{\bf (CT8) Complicated Causal Pathways in Brain Disorders.}  
Brain disorders (e.g., AD) are affecting 1 in 6 people worldwide and pose a massive threat
to public health, resulting in significant disability, morbidity, and mortality.   Most approved therapies for treating brain disorders only treat symptoms. Existing  studies suggest that  most complex brain disorders are   highly heritable with polygenic architecture and are caused by a combination of genetic and health 
factors \cite{miller2016multimodal,alnaes2019brain,van2016genetic,zhao2016annual}.  
Moreover, many brain disorders can be regarded as endpoints of abnormal trajectories of brain
  changes.    Since    neuroimaging measures are closer to the underlying biology and can be measured temporally,  much effort has been devoted to  understanding the temporal CGIC pathophysiological pathway in the continuum of brain disease progression  from increasingly large cohorts (e.g., ADNI). It may  lead to the identification of possible hundreds of risk genes and  
health  factors that contribute to abnormal developmental trajectories of brain disorders. 
 Once such
identification has been accomplished, 
we may establish a set of complex  causal relationships that delineate the CGIC pathways confounded with environmental factors and unobserved confounders  as shown in Figure \ref{fig:integration}. 
These risk trajectories can be detected early enough to identify urgently
needed therapies that target the correction of abnormal developmental trajectories, ultimately preventing
the onset of brain disorders and reducing their severity. 

\subsection{PSA Methods}

There is a great interest in developing various SL methods for NDA in order to address (CT1)-(CT4) inherent in neuroimaging data discussed in Section 3 and  (CT5)-(CT8) in large-scale neuroimaging  studies discussed in Section 4.  We briefly review nine categories of  PSA methods in the  literature, many of which are emerging. Moreover, we cannot cite many important papers in each category due to the maximum number of references  set by the publisher.

 \subsubsection{Study Designs}
 Popular designs in large-scale observational studies include case-control, cross-sectional, and cohort studies   \citep{thompson2012sampling,riffenburgh2012statistics}. 
 These designs  can be applied to a variety of scientific questions, but they all have certain limitations when it comes to specific  clinical and epidemiological applications. 
Case-control studies are good for studying rare clinical outcomes and  latent diseases.  
 Participants in a case-control study are selected based on their outcome status and are defined as cases and controls. 
 In such studies, matching is often used to ensure that the cases and controls have similar characteristics (such as age and sex), which can increase study efficiency. 
 Wellcome Trust Case Control Consortium, for example, uses a case-control design in order to study multiple major diseases with the careful use of a common control group \citep{wellcome2007genome}. 
 The case-control design has been widely combined with meta-analysis approaches to pool summary-level data from different research groups, such as the Psychiatric Genomics Consortium \citep{watson2020psychiatric} and ENIGMA \citep{thompson2020enigma}.  However, the selection and matching steps may be prone to certain biases and confounding effects, such as selection bias and recall bias. Due to potential differences between study samples and the general population, the findings and statistics learned from case-control designs may not have perfect generalizability.
As neuroimaging data were frequently collected as secondary traits or endophenotypes in these biomedical studies, the "case-control" nature needs to be taken into account when inferring these imaging traits in statistical analyses. 
 
On the other hand, cohort studies recruit participants without screening for the outcome of interest.
Participants are selected based on their characteristics and/or their willingness to volunteer. 
The outcome of interest is typically monitored over time to assess its occurrence and the relationship between outcome and exposures can be evaluated at  baseline (e.g., cross-sectional analyses) or in a longitudinal framework. For example, the UKB is a large, population-based cohort study \citep{littlejohns2020uk,miller2016multimodal}, and many cross-sectional analyses have been conducted based on baseline data from  UKB. 
However, UKB is well known for its "healthy volunteer" selection bias, and may not be a true representation of the general population \citep{fry2017comparison}. 
To deal with selection bias, reweighting-based methods could be used from a causal inference perspective \citep{ batty2019generalisability,bradley2022addressing}. 
These methods typically 
assume that volunteer bias can be explained by observed variables, such as socioeconomic status. In addition, missing data is also a known source of confounding in cohort studies, especially when the outcome of interest is not independent of the missing mechanism.  Failing to address these biases may lead to confounding effects, biased statistical results, and  misleading findings.

Moreover, when meta- or mega- analyses integrate data from different studies and cohorts, the study designs of these sources may differ. 
Ignoring such differences may lead to unexpected results in data integration. For example, it may not be straightforward to specify a correct statistical inference framework when pooling data from a case-control and a cohort studies. It is obvious that naive analyses without taking into account of the study design will lead to biased findings. Therefore, it is important to understand sampling mechanisms and to apply them appropriately for the desired objectives when designing and merging population-based studies.  
 
As compared to observational studies, there are fewer experimental studies in population-based biomedical research.  One of the reasons is that it is typically difficult and expensive to conduct experiments on a large number of subjects. However, experiments play a key role in advancing our understanding in biomedical data science. For example, well-designed task/event-based fMRI experiments can help understand the brain functional changes due to human behavior and interventions. In addition, sequential decision making is also important to better design of the follow-up stages in a large-scale population-based study.  In summary, the sampling mechanism needs be taken into consideration when interpreting and generalizing findings from observational studies. It is evident that large-scale experimental designs for NDA are seriously lacking in major publicly available data resources, and this issue will require greater attention in future biomedical data science research.

  \subsubsection{Statistical Parametric Mapping (SPM)}
  
 There is a large literature on the development of  various statistical methods, called Statistical Parametric Mapping (SPM), 
 for two major  NDA tasks  including  image reconstruction from  image volumes within each subject and group analysis of  images obtained from different subjects/groups. In both tasks, images are assumed to be registered to the same space. 
 We will briefly review conventional SPMs and their extensions below.

 The SPM 
  refers a statistical technique for detecting  changes in  brain structure and function recorded during  neuroimaging experiments within individual subjects or across groups.  Such SPM has been  implemented in popular 
neuroimaging software platforms
  including statistical parametric mapping (SPM) (www.fil.ion.ucl.ac.uk$/$spm$/$) and  FMRIB Software Library (FSL) (www.fmrib.ox.ac.uk$/$fsl$/$).  It consists of three key modules: (i) smoothing neuroimaging data  spatially and/or temporally, 
  (ii) fitting voxel-wise general linear models, 
  and (iii) correcting for multiple comparisons by using random field theory (RFT), false discovery rate (FDR), and permutation method.  
 Despite the popularity of  SPM, there is a great need to extend it in three important directions.

 The first direction is to address  several major drawbacks of  Gaussian smoothing method, which may  dramatically increase  
the numbers of false positives and false negatives \citep{zhu2014spatially}. 
Moreover, for   twin studies, \cite{li2012twinmarm} showed that smoothing raw images can dramatically 
 decrease statistical power in detecting environmental and genetic effects, which is critically important for imaging genetic studies.  
To address those drawbacks,   multiscale adaptive models have been proposed to 
extend  the propagation-separation method   to a large class of parametric and semiparametric models for group analysis  \citep{polzehl2010structural,zhu2014spatially,Zhu:2011,li2012twinmarm}. Those multiscale adaptive methods  dramatically  increase signal-to-noise ratio, while preserving  spatial details.

The second direction is to move from general linear models (GLMs) to more advanced statistical models. Such development is primarily motivated by dealing with complex study design,  sampling bias, missing data, complex data objects, and complex relationships as discussed in (CT4)-(CT8).
Simply applying general linear models to all scenarios in (CT4)-(CT8) can easily lead to false positive and false negative results.   
In the era of large-scale neuroimaging studies, it is important to integrate and extend many statistical packages in  professional statistical softwares including R (www.r-project.org$/$),
RStudio (www.rstudio.com), 
SAS (www.sas.com), and python statsmodels
(www.statsmodels.org), among others. 
It opens a new world with many parametric, semiparametric, and nonparametric statistical models and their associated statistical inference tools, even though they may not directly applicable to NDA without some modifications.

There are two ways of applying and extending those statistical models in statistical softwares. The first one is to apply those statistical models to neuroimaging data, generate statistical maps for various  statistical results (e.g., $p-$values, parameter estimates, and diagnosis measures) across   spatial locations (e.g.,  voxels, vertexes, or pixels), and then perform multiple comparisons. 
We will discuss how to correct for multiple comparisons in details below.  
Minimum effort is required for all necessary technical developments.  
The second one is to explicitly incorporate the spatio-temporal structure discussed in (CT2) into different statistical models and then  correct for multiple comparisons. For instance,  some notable developments include 
multiscale adaptive regression methods for longitudinal neuroimaging data
\citep{yuan2014fmem}, spatial varying coefficient models  \citep{zhu2012multivariate,li2021sparse, 
zhu2014spatially,zhang2020image}, quantile models  \citep{zhang2021high,yang2020quantile}, and   
functional principal component analysis (fPCA) 
\citep{chen2019functional}, among others.  

We want to make four remarks on different statistical models for modeling neuroimaging data. 
First, most statistical models for SPM can be regarded as an approximation to model (\ref{eq1}) in order to disentangle the signals of interest, such as age, gender, or diagnosis. 
Second, most statistical models for SPM can be formulated as an image deconvolution problem according to equation (\ref{eq2}).  
Third, although quantile methods have not been widely used in NDA, they improve our   
 understanding of the  conditional distribution  of imaging measures on the spatial domain that may have nonlinear relationships with  various predictors in model (\ref{eq1}). 
Fourth, it should emphasize that most functional data analysis (FDA) methods in statistics were  primarily developed for one-dimensional curves 
\citep{silverman2005functional,wang2016functional} and extending these FDA methods to 2 and higher dimensional neuroimaging data faces major statistical and computational challenges.

 The third direction is to develop statistical methods, including RFT,  resampling methods, and FDR,  to correct for multiple comparisons in NDA.   Most RFT and resampling methods  
 control  for the familywise error rate
 by accounting for the spatio-temporal structure of raw neuroimaging data as discussed in (CT2), whereas   
most FDR methods directly operate on uncorrected $p-$values without addressing (CT2). However, 
recently,  several  FDR methods have been developed to  control for FDR  in multiple testing of spatial signals \citep{sun2015false,zhang2011multiple}. 
Although FDR is  applicable to a larger class of statistical models beyond GLMs, it does depend on the computation of uncorrected $p-$values, which is nontrivial in many cases.

 Since the beginning of fMRI, RFT dominates the field of NDA primarily due to many fundamental contributions made by Drs. Worsley, Adler, Nichols, Taylor and their collaborators \citep{worsley2004unified,adler2007random,nichols2003controlling}. 
  RFT has been widely used for voxelwise  and cluster size inference in order to test for the intensity of an activation and for the significance of the spatial extent of an activation.  
 Voxelwise RFT uses the expected Euler characteristic heuristic of random fields to approximate the  $p-$value of the maximum statistic, whereas 
 cluster-size RFT uses the distribution of  the maximum of cluster sizes in
a zero mean stationary random field. However, 
current RFT results cannot meet important requisites for  many advanced statistical models in NDA due to two primary reasons.  
First, 
most RFT results are limited to GLMs and some minor extensions \cite{adler2007random}. It requires substantial effort on the development of new RFT results for more advanced models. 
Second, 
most RFT results require strong assumptions including 
  stationarity and high order smoothness,  that are often invalid for fMRI. 
  Specifically, \cite{eklund2016cluster} had two important observations: (i) some key  assumptions of RFT are invalid for fMRI, and  
  (ii)  the existing RFT  can lead to inflated false positive rates for cluster size inference.

Resampling methods primarily include permutation and bootstrap-based methods, both of which approximate the null distribution of test statistics conditional on the observed data. Although permutation testing has received some attention in NDA, it has not gained much attention in statistics lately due to computational and methodological challenges. Specifically,  permutation methods require complete exchangeability under the null hypothesis, which can be problematic even for the simplest two group comparison problem.  Bootstrap-based methods, particularly wild boostrap, have gained substantial attention in statistics due to their flexibility, theoretical ground,  and good empirical performance, even though additional effort may be required for further development and application of good wild bootstrap methods to different models.
Theoretically, resampling methods like wild boostrap are shown to be valid conditional on data \citep{kosorok2003bootstraps,chatterjee2005generalized}. 
Practically, wild bootstrap methods have been successfully applied to NDA, including 
a heteroscedastic linear model for surface analysis \citep{Zhu2007a}, 
regression analysis of asynchronous longitudinal functional and scalar data
\citep{li2020regression},  
functional mixed models for longitudinal neuroimaging data
\citep{yuan2014fmem},   
and statistical models for imaging genetics \citep{Huang2015,huang2017fgwas}.

As an illustration, we consider an interesting study \citep{botvinik2020variability}, that examined the  
variability of different SPM analytical pipelines in  the analysis of a single neuroimaging dataset by 70 independent teams.
Sizeable variations in the final statistical results of hypothesis tests are caused by all three modules of SPM. 
A surprising observation is that the 
 spatial smoothness of fMRI is the strongest factor in explaining such variation. 
 Another study further evaluated the effect of different fMRI preprocessing pipelines on analytical results \citep{bowring2019exploring}.
 Both studies  call for additional development of resources and methods for  the reduction of variability in preprocessing and analysis pipelines and the effect of the variability on analytical results.

   \subsubsection{Object Oriented Data (OOD) Analysis}

    We will briefly review OOD and their extensions below.
Object oriented data (OOD) analysis is 
a comprehensive statistical framework   including estimation methods and statistical 
  theory for the analysis of populations of complex objects \citep{marron2021object,huckemann2021data,srivastava2016functional,wang2016functional,Dryden1998}. Some specific examples of complex objects given in (CT7) can be  
  elements of mildly non-Euclidean spaces, such as Riemannian symmetric spaces, or of strongly non-Euclidean spaces, such as spaces of tree-structured  objects.   A primary application of OOD in NDA is  group analysis of  complex objects extracted from neuroimaging data.

    Three classes of analytical procedures for OOD include (i) feature analysis,  (ii) extrinsic analysis, and (iii) intrinsic analysis.
    The key ideas of feature analysis include using some feature extraction functions to project random objects to Euclidean-valued variables and then applying the second and third modules of SPM to those Euclidean-valued variables. A key advantage of the feature analysis is its computational efficiency. Moreover, Euclidean-valued variables projected from random objects can be biologically meaningful, if their corresponding extraction functions have strong biological interpretation. We consider two examples of feature analysis as follows. The first one is to treat diffusion tensors, which are $3\times 3$ symmetric positive definitive (SPD) matrices, as random objects. It is common to  calculate several invariant measures of diffusion tensor, such as fractional anisotropy (FA), and then use SPMs to analyze FA images. 
    In neuroscience, FA is an indirect measure of fiber density, axonal diameter, and myelination in white matter.  
    The second one is to treat functional brain network as random objects and use the feature analysis to understand the topological organization of brain networks.  Specifically, one may calculate  various graph metrics (e.g., nodal centrality, network efficiency, or degree) of functional brain network and then perform  the group analysis of  these graph metrics \citep{bullmore2009complex,simpson2013analyzing}. For instance, in brain network, 
    network efficiency   describes how brain network efficiently exchanges information.  
    However, it is often nontrivial to develop a good feature extraction function with strong neuroscience interpretation besides 
    having the feature vector contain  partial information about the original object.

    The key ideas of extrinsic analysis are (i) to 
    embed  the curved space where the object
resides onto some higher-dimensional Euclidean spaces, (ii) to perform statistical inference on random objects in the embedded Euclidean space, and (iii) to pull results back onto the curved space.  
 A key advantage of extrinsic analysis is its computational efficiency. Existing extrinsic analysis methods have been developed for mean,   median,   local regression,  and  dimension reduction \citep{lin2017extrinsic}.  For instance, 
 diffusion tensors can be embedded in a 6 dimensional Euclidean space, whereas the $d-$dimensional sphere $S^d$ can be embedded in the $(d+1)-$dimensional Euclidean space.  The manifolds
considered in directional statistics are spheres and projective
spaces and the associated statistical tools are primarily 
extrinsic approaches.
However, there are two drawbacks. First, it is nontrivial to propose a good equivariant embedding in most cases, which requires substantially deep thinking. Specifically, in step (i),   equivariant embeddings are required to  preserve  a lot of  geometry of  the original curved space. Second, in many cases, it is unclear as to how one could pull results back onto the curved space.

 The key ideas of intrinsic analysis are (i) to 
    introduce a 'good' metric $\rho$ for the curved space $\mathcal M$ where the object
resides, denoted as $({\mathcal M}, \rho)$, and (ii) to perform statistical inference on random objects in $({\mathcal M}, \rho)$.  
Some examples of metric spaces with additional structure include Riemannian manifolds, normed vector spaces, length spaces, and graphs. 
For instance,  a Riemannian manifold $(\mathcal M, g)$ is a real, smooth manifold $\mathcal M$ equipped with a Riemannian metric tensor $g$ defined for all tangent vectors at every point.   One can define  the geodesic distance between two points  on a connected Riemannian manifold.  
We can further construct quotient metric spaces for $({\mathcal M}, \rho)$ based on an equivalence relation on $\mathcal M$, denoted as 
$\sim$ by 
endowing the quotient set ${\mathcal M}/\sim$ with a pseudometric  $\rho_{P}$. 

A fundamental issue in intrinsic analysis is how to appropriately introduce a good metric $\rho$ for 
$(\mathcal M, \rho)$ or metric tensor $g$ for $(\mathcal M, g)$.  The choice of $\rho$ (or $g$) has fundamental effects on 
downstream computation and statistical inference. For instance, \cite{dryden2009non} discussed eight different metrics of the space of SPD for estimation of mean diffusion tensor.  Recently, \cite{srivastava2016functional} introduced a general elastic metric, which includes the Fisher-Rao metric as a special case, for the shape analysis of curves, allowing us to separate phase and amplitude components. 
In general, the choice of $\rho$ (or $g$) should  focus on  the signal of interest and data variability  in random objects, while 
considering  computational efficiency.  

In the last decade, significant progress has been achieved  in the development of intrinsic statistical models for manifold-valued data in finite-dimensional Riemannian manifolds.   Frechet mean, median, and variance provide a simple way of characterizing center and variability of random objects in $\mathcal M$ \citep{arnaudon2013medians,marron2021object,huckemann2021data}. Principal geodesic analysis \citep{fletcher2004principal} is further developed to reduce the dimensionality of random objects, while increasing interpretability and minimizing information loss. \cite{cornea2017regression} developed an intrinsic regression model based on Riemmannian logarithm and exponential maps  for random objects in a Riemannian symmetric space. Other notable contributions include 
Riemannian functional data analysis, intrinsic local polynomial regression, Wasserstein regression, a generic measure of dependence, and longitudinal analysis 
\citep{yuan2012local,shao2022intrinsic,chen2021wasserstein,pan2019ball}, among others. 
Despite these new developments, computing  intrinsic estimators is notoriously difficult, requiring more attention. 

Statistical shape modeling and analysis  have emerged as important
tools for understanding brain structure and function extracted from
neuroimaging data.  
Four key components of  shape analysis include (i) shape representation, (ii) shape distance between shapes, (iii) shape registration, and  (iv) group analysis of shapes. Shape  analysis methods  depend on shape representations  including   landmarks, implicit representations, parametric representations, medial models, and deformation-based descriptors, among others   \citep{marron2021object,Miller2009,Grenander2007,srivastava2016functional,Dryden1998,chung2007weighted,fischl2012freesurfer}.   
Most earlier representations focus on either points on the object boundary or parametric descriptors of the object boundary, whereas deformation-based representations use shape information in the entire image.  
Most shape spaces
are quotient metric spaces  based on an equivalence relation including  
translation, rotation, and scaling.   
 Some notable shape analysis methods include the large deformation diffeomorphic metric mapping (LDDMM) technique \citep{Grenander2007}, 
the elastic statistical shape analysis \citep{srivastava2016functional,ZhangZhu2023}, and the Wasserstein shape analysis 
\citep{shi2019hyperbolic}.

  \subsubsection{Imputation Methods} \label{imputation}

Developing good imputation methods for neuroimaging data requires  a deep knowledge and understanding of  
reasons for missing data  and their mechanisms in NDA.
 Table \ref{tab: missing}  summarizes some common reasons for missing data and their corresponding missing mechanisms in NDA.
Reasons for missing data  in NDA include  missing image modalities due to different acquisition protocols, different study designs, data transfer and storage loss, faulty scanning due to image corruption and susceptibility artifacts, and  participant attrition due to allergies to materials, personal belief, and financial costs, among others.   There are three missing mechanism categories including MCAR, MAR,  and MNAR \citep{LittleRubin2002,ibrahim2009missing}. Distinguishing between MAR and MNAR depends on whether the missingness is  predictable based on either observed covariates  or  missing variable itself. For example,  if dropout rates differ  according to observed covariates (e.g., age, sex, or race), then the missing mechanism is traceable and MAR. In contrast,  if dropout depends on missing data itself,  then it is MNAR and ignoring such missingness may introduce substantial bias.  
MCAR, as a special case of  MAR, assumes that the distribution of the missing data is indistinguishable from the non-missing data. Such assumption is strong and usually difficult to meet in  practice. 
In general,  when values are missing systematically,   downstream  data analysis without correcting for missing data may lead to erroneous conclusions.


\begin{table}[!h]
\begin{center}
\footnotesize
\begin{tabular}{ |p{1.8cm}||p{3cm}|p{6cm}|  }
 \hline
Missing mechanism&Causes& Details\\
 \hline
\multirow{2}{*}{MCAR}  &Faulty scanning & Removal of images with corruption or susceptibility artifacts  \\ \cline{2-3}
  &Faulty scanning & Random failure of experimental instrument   \\ \cline{2-3}
  &Data loss & Data transfer/storage loss  \\ \cline{2-3}
  &Data loss & Missing entries  \\ \cline{2-3}
  &Attrition/Nonresponse & Unable to participate due to migration/move (irrelavant with the study)  \\ \cline{2-3}
  &Study design & Study ended early  \\ \cline{2-3}
   &Study design & Modalities were not included in the imaging protocol  \\ \hline
 \multirow{2}{*}{MAR} &Study design & Exclusion criteria, such as age, sex, race, socioecnomic status, etc. \\ \cline{2-3} 
 &Attrition/Nonresponse & Dropout due to side effects, such as allergy  \\ \cline{2-3} 
  &Attrition/Nonresponse & Dropout rates vary among different age or sex groups \\ \hline
  \multirow{2}{*}{MNAR} &Study design & Quit the study due to physical or psychological health conditions \\ \cline{2-3} 
 &Attrition/Nonresponse & Dropout due to concerns of financial cost  \\ \cline{2-3} 
 &Attrition/Nonresponse & Dropout due to concerns of limited available time to visit  \\ \cline{2-3} 
 &Attrition/Nonresponse & Dropout due to concerns of scanning safety  \\ \cline{2-3} 
 &Attrition/Nonresponse & Dropout due to concerns of personal data unauthorised disclosure \\ \cline{2-3} 
  &Attrition/Nonresponse & Quit the study, following another person's behavior \\ \cline{2-3} 
  &Attrition/Nonresponse & Deliberately not willing to respond \\ \cline{2-3} 
 \hline
\end{tabular}  
\end{center}\caption{A summary of  scenarios with different missing mechanisms in cognition/behavior-related studies}\label{tab: missing}
\end{table}

There are at least two main  strategies for  handling missing data including omission and imputation \citep{nakagawa2011model,LittleRubin2002,ibrahim2009missing}. 
Common omission approaches include listwise/pairwise omission and dropping features. 
 Although omission
is simple and easily used,  it can lead to serious estimation bias, large efficiency loss, and dramatic reduction of statistical power. 
There are two types of imputation methods, including single imputation and multiple imputation. 
Single imputation methods generate  one imputation value for each  missing observation, which leads to a single complete data,  while treating the imputed values as the true values in downstream data analysis.  Therefore, downstream analyses based on the single imputed complete dataset do not account for the imputation uncertainty. 
Two main strategies of single imputation including imputation by  statistical values (e.g., mean, median, or maximum) and imputation by predicted values generated from a statistical model. 
Multiple imputation methods generate
many imputed values for each missing observation,  which lead to many complete datasets,  while 
 analyzing all of them in downstream data analyses.    The use of multiple imputation allows us to explicitly account for  imputation uncertainty.

Some additional  statistical challenges arise from  handling missing neuroimage data due to (CT1)-(CT4), even though 
 both omission and imputation methods are useful methods for NDA.  Specifically,     
as discussed in Section 4 and Figure 4, image data  are largely block-wise missing, 
while there are a large number of features across different domains (e.g., genetics/genomics)  in various biomedical studies.   
In this case, it requires building image imputation  models to impute missing high-dimensional images  conditional on all other observed features, which may include  other imaging modalities, genetic/genomics, and demographic variables.   
A promising research topic is to develop deep generative models, which have been used to achieve impressive results in   image generation and image-to-image  translation  for image imputation models.  
In particular,   image-to-image translation is designed to learn the mapping between an input image and an output image, while preserving the content representation \citep{alotaibi2020deep}. This task can be further classified into paired and unpaired imputation according to whether both input and output images are available on the same subjects in the training data. For instance, 
conditional generative    adversarial network (CGAN) methods, such as the Pix2pix \citep{isola2017image} 
method, perform pixel-to-pixel image synthesis using paired image data, whereas CycleGAN \citep{zhu2017unpaired} was developed to model image translation based on unpaired data.  
Although there has been development of many image-to-image translation models for specific neuroimaging pairs, these models   
  require  substantial effort in validation requiring the use of synthetic and real data sets for downstream tasks, such as prediction.   
Furthermore, it is interesting to incorporate additional information (e.g., genetics, diagnosis status, and sex) to impute missing image data, while imposing their dynamic causal relationships in Figure \ref{fig:integration}. However,   
   little has been ever done on the development of  CGAN-based imputation models for neuroimaging data along this direction. 
In addition,  since image data may be missing under MNAR as detailed in Table 1, it is important to develop CGAN imputation models under MNAR. 
  \subsubsection{Data Integration (DI)} \label{DIDI}

  We witness the exponential increase in the collection and availability of multi-view data, including electronic health records, imaging, genetic, sensor data,  and text,  from different studies and clinics as discussed in Section 4. 
   Data integration (DI) is the process of  integrating   multi-view  data from different sources  into  a unified view of information for better data management and downstream tasks.  
     A good DI system consists of  (i) a feature engineering pipeline for generating  more complete 'high-quality' data and their associated  features,  (ii) SL methods for data integration associated with different NDA tasks, and (iii) a feedback loop to   improve data collection and feature extraction for major NDA tasks.      
 The feature engineering pipeline consists of  data ingestion, data processing, data annotation,  transformation, and storage.
Missing data imputation prevails in all these tasks, but we defer its related methods to Subsection \ref{imputation}.
However, although much progress has been achieved in the last decade,  it remains   challenging to develop 
  a good DI system for  NDA  due to the fact that they 
   are complex, heterogeneous, temporally dependent, irregular,  poorly annotated,  and generally unstructured as discussed in (CT2)-(CT8).

We  review  SL methods for data integration within individual studies  and  across studies associated with four major NDA tasks including 
(T1) multimodel neuroimaging fusion, (T2) the genetic architecture of neuroimaging measures, (T3) gene–environment interaction on neuroimaging measures, and 
(T4) the GIC pathways. 
 We defer 
 most SL methods for (T2) and (T3) to Subsection \ref{imagingGenetic} and those for (T4) to Subsection \ref{causal}.  
Popular building blocks in  SL methods for data integration include  feature concatenation, Bayesian models, tree-based ensemble methods, multiple kernel learning,  matrix/tensor factorizations, and DL \citep{li2018review,zhao2017multi}. 
For instance, Bayesian methods can easily incorporate prior information from different views, whereas tree-based methods can use ensemble methods to integrate trees learned from each view.

As an illustration, we consider matrix factorizations and DL for data integration in a single study. 
First, we consider 
  a generic model for using matrix factorizations for multi-view integration in a single  study.  
Suppose that we observe a $p_k\times n$ row-mean centered data matrix, denoted as $I_k$,  
for the $k-$th view of $K$ views on $n$ subjects, where $p_k$ is the number of variables and $n$ is the number of subjects. 
A generic model for  matrix/tensor factorizations is given by 
\begin{equation}\label{eq3}
I_k=C_k+D_k+E_k~~~~~\mbox{for}~~~~~ k=1, \ldots, K, 
\end{equation}
where  $C_k$ is a low-rank
common-source matrix representing  latent factors common across all views, $D_k$ is a low-rank distinctive-source matrix representing distinctive latent factors of the corresponding view, and $E_k$ is the noise matrix.
Some state-of-the-art matrix factorization methods based on model (\ref{eq3}) include  
common orthogonal basis extraction \citep{zhou2015group}, joint and individual variation explained \citep{lock2013joint}, and  
decomposition-based generalized canonical
correlation analysis \citep{shu2022d}.  They differ from each other in how to reconstruct the common- and distinctive-source matrices. 
Second, we consider  the hierarchical architecture of DL for multi-view integration as another powerful method. Its hierarchical structure consists of (i) the construction of    sub-networks  $s_k={\mathcal N}_k(I_k)$ (e.g., 
 Variational Autoencoder (VAE) and  Generative Adversarial Network (GAN)  for neuroimaging data)  for  $k=1, \ldots, K$   and  (ii) the integration of all individual sub-networks into a model $Y=f(s_1, \cdots, s_K; \theta)+\epsilon$, where $f(\cdot)$ is a link function, $\theta$ is a vector of parameters,  and $\epsilon$ is an error term. We can use an objective function similar to equation (\ref{eq2}) to tune $\theta$ and $\{\mathcal N_k\}$.   
 \cite{miotto2018deep} discussed different architectures of  sub-networks for individual views. These sub-networks can be first adopted from some pretrained models from other fields, such as computer vision,  
 and then be tuned in the whole model at the integration stage.

We consider two major methods  for data integration across multiple studies or centers including  {\it   the merged learner}  and {\it   the ensemble learner}.  
The {\it   merged learner} proceeds with merging and processing  data from all studies  and then training a single learner based on the merged data. 
It is common to use  fixed- or random-effect models  to train the learner \citep{zugman2022mega}. 
The {\it  ensemble learner} proceeds with training a learner based on the data obtained from each study and then using 
	a weighted average of all learners. It includes ensemble machine learning \citep{Guan20}, meta-analysis \citep{meta_random},  fusion learning  \citep{fusion2020},  and federation learning 
	\citep{li2020federated}, among others. 
ENIGMA has been using  the ensemble learner  in most of their imaging genetic studies,  but 
  it starts to use  the merged learner (or mega analysis)  \citep{zugman2022mega}. Since 
 data pooling can  dramatically increase sample size and  ensure consistent data processing and quality control, 
  the merged learner will be taken  in more and more international neuroimaging efforts.   
  
 We discuss two major issues in mega analysis including  heterogeneity discussed in (CT4) and  sampling bias in (CT5). 
 First, there is a great interest in developing data harmonization methods to explicitly correct additive site and scanner effects,     covariance batch effects, hidden factors, as well as 
	  some structural priors in    neuroimaging data   \citep{yu2018statistical,chen2021privacy,HuangZhu2022}. These methods 
	  partially remove the effects of  those confounding variables of not interest, but they require extensive validations by using walking phantoms,  synthetic datasets, and annotated data sets. Second, although it is tempting to pool multi-view data from studies with 
	  different study designs, simple statistical methods based on fixed  and random effect models 
	  \citep{burke2017meta,simmonds2015decade} cannot appropriately handle such issue. We discuss several key problems. First, in many imaging related studies (e.g., ADNI and UKB), neuroimaging data are only the secondary phenotypic variables, so it can be very problematic not to adjust sampling bias even in a single study \cite{kim2015cautionary,zhu2017genome}. Second, many neuroimaging-related studies 
	  have different study designs and may have minimum overlap in some key confounding variables (e.g., age) of interest. For instance, besides their age differences,  there are many twins in HCP, whereas ADNI has many longitudinal observations. It raises 
	  many serious issues on the target population for the merged sample, the type of scientific questions to be answered, and the choices of different statistical models (e.g., prospective and retrospective likelihood).    In conclusion, one cannot simply perform the merged learner for many NDA tasks without appropriately addressing sampling bias in (CT5).

     \subsubsection{Dimensional Reduction (DR) Methods}
  
    The goal of DR is to transform  data  from a high-dimensional space to a relatively low-dimensional space, while retaining important information of the original data. There is a large literature on the development of  various statistical methods  
 for dimension reduction (DR) due to (CT3). 
 We can cluster DR methods into  feature selection and feature extraction.   
Feature selection aims  to find a subset of the original features for a specific task, whereas  feature extraction aims to construct new features from the original features.
Originally, the aforementioned DR methods were developed to solve the {\it small-$n$-large-$p$} problem, where the number of subjects is much smaller than the number of imaging variables. 
However, with the availability of many large-scale neuroimaging studies as discussed in Section 4,  we have to deal with the {\it large-$n$-large-$p$ } problem, in which both the number of subjects and and the number of variables  are both extremely large. This large-$n$-large-$p$ problem requires further development in DR methods.

The feature selection methods  can be further grouped into the filter strategy, the wrapper strategy, and the embedded strategy based on how 
the selection algorithm and the model building are combined \citep{li2017feature}.   
Filter  methods use a selection measure, such as correlation and distance correlation, to select a feature subset. 
 Wrapper methods, such as stepwise regression, use a search algorithm  based on a predictive model to score feature subsets.
 Embedded methods, such as decision tree and LASSO,  select features as part of the model construction process. 
 In practice,  feature selection is essential to  eliminate  a large number of noisy variables before running downstream data analysis.

The feature extraction methods  can be categorized into knowledge-based and  data-driven approaches. 
In NDA, the knowledge-based feature extraction  is to use specific human brain atlases to perform feature extraction within individual regions and across region pairs. The use of several tens to several hundreds of homogeneous regions of interest (ROIs) in brain atlases  dramatically reduces the complexity of multiple neuroimaging data.  
  It improves neuroanatomical precision for studying  the structural and functional organization  of the human brain. 
 The data-driven feature extraction methods can be grouped into unsupervised, supervised and semi-supervised approaches for both traditional approaches and modern DL, respectively \citep{anowar2021conceptual,liu2021self}.  Some notable examples of unsupervised feature extraction methods  include  principal component analysis (PCA), kernel PCA (KPCA), functional PCA (FPCA),  single value decomposition (SVD),  tensor decomposition,   multidimensional scaling (MDS),   and independent component analysis (ICA).  
 See \cite{anowar2021conceptual}   and references therein for a systematic review and empirical comparisons of 
various  unsupervised DR approaches.  
Some notable examples of supervised DR methods include 
linear discriminant analysis (LDA), partial least squares regression (PLSR),   and canonical correlation analysis (CCA). 
Feature extraction and feature selection methods have been integrated together to solve the small-$n$-large-$p$ problem, while accounting for 
complex spatio-temporal structures in (CT2) \citep{lin2014correspondence,zhu2017mwpcr}.
However, while most existing feature extraction methods are infeasible for   the large-$n$-large-$p$ problem due to  limited computing speed and computer memory, several hierarchical feature extraction methods have been developed to address related challenges 
\citep{crainiceanu2011population,gong2021phenotype}.

  Most unsupervised DL approaches (or the self-supervised learning (SSL)) to extract image embeddings   includes three classes: the generative, contrastive, and adversarial approaches \citep{liu2021self}. These SSL approaches train the encode-decoder networks by encoding  input images into a low-dimensional representation,   to contrasting semantically similar and dissimilar pairs of embeddings, and generating fake samples that a discriminator can hardly distinguish from real samples. Recently,    semi-supervised SSL (SS-SSL) have been developed by incorporating downstream tasks, such as classification or prediction, to original pretext tasks (construction and contrasting) \citep{jaiswal2020survey}. Comparing  with traditional DR approaches, DL-based DR approaches usually extract more informative representations by taking advantage of great computing power and more flexible frameworks.

  \subsubsection{Imaging Genetics} \label{imagingGenetic} 
   {
   The genetic architectures of human brain structures and functions are of great interest. Using imaging traits as phenotypes, the extent to which genetics can affect the structure and function of the human brain (or, heritability) has been quantified in previous family or population-based studies \citep{blokland2012genetic,zhao2019heritability}. 
   Several consortiums, such as the the ENIGMA \citep{thompson2020enigma},
   the CHARGE \citep{fornage2011genome}, and the IMAGEN \citep{mascarell2020imagen},  were established to discover the genetic loci associated with human brain structure. In recent years, large-scale MRI datasets, such as UKB and ABCD, have provided further insights into the genetic determinants of the human brain. For example, \cite{elliott2018genome} and \cite{smith2021expanded} screened more than 3,000 brain functional and structural imaging phenotypes from the UKB study. The genetic architecture of commonly used imaging traits, such as the regional grey matter volumes from sMRI \citep{zhao2019genome}, WM microstructure from DWI \citep{ZhaoZhu2021uk}, and functional connectivity from fMRI \citep{zhao2022fMRI} have been discovered. From these studies, hundreds of brain-related genetic loci have been identified and substantial genetic overlaps with major brain disorders were observed, such as AD and schizophrenia. Several open resource knowledge portals have been developed in imaging genetics, including the Oxford BIG40 (\url{https://open.win.ox.ac.uk/ukbiobank/big40/}) and BIG-KP (\url{https://bigkp.org/}). While they extract imaging features using distinct pipelines, they provide similar findings regarding the genetic control of the human brain. Figure \ref{fig: icc}B presents the heritability values of various imaging phenotypes based on UKB. 
   
A typical imaging genome-wide association study (GWAS)  contains the following steps. First, we develop and/or apply imaging data analysis pipelines to extract imaging features from raw neuroimaging data. For example, in the WM GWAS \citep{ZhaoZhu2021uk}, we applied the ENIGMA-DTI pipeline to extract WM microstructure measures from over 40,000 subjects \citep{jahanshad2013multi}.
Although voxel-wise or vertex-wise feature maps are available, aggregate measures at the brain region-level imaging traits (such as ROI and WM tracts) are typically used in subsequent genetic discoveries. In addition to improving the signal-to-noise ratio, these region-level traits may reduce the burden of multiple testing, while increasing the statistical power in genetic analysis. Second, variant-level and gene-level association analysis are performed to detect significant genetic variants or genes in a large-scale discovery cohort. An independent holdout cohort, which is typically smaller than the discovery one, will be used to examine if the significant trait-variant/gene associations can be replicated.  Further replications and generalizability can be explored using racially diverse cohorts. 
Additionally, polygenic risk scores can also provide evidence of validation by evaluating the proportion of variance of imaging traits that can be predicted by genetic variants.

A few tools have been developed to estimate the heritability using individual-level 
(e.g., GCTA-GREML  \citep{yang2011gcta}) or summary-level data (e.g., univariate LDSC \citep{bulik2015ld}). Furthermore, 
partitioned LDSC  can be used to estimate the  enrichment of heritability related to specific brain tissue or cell types, such as glia and neurons. FUMA \citep{watanabe2017functional} is a useful platform for functional gene mappings based on summary-level data. Coloc, bivariate LDSC, and Mendelian randomization methods \citep{sanderson2022mendelian} can quantify the genetic relationships between imaging traits and  other complex traits or diseases from different perspectives.  See 
 \cite{sun2020statistical} for a  recent review of GWAS methods.

  Despite recent significant advancements in imaging genetics, it remains challenging  to map the causal biological pathways linking genetics and brain abnormalities to neuropsychiatric disorders \citep{le2019mapping,shen2019brain}. See Figure \ref{fig:integration} for a hypothetical causal pathway. 
   To understand the causal pathway, by which genetic variation impacts risk for brain diseases, neuroimaging can serve as important endophenotypes. The identified genetic loci in large-scale imaging genetic cohorts need to be integrated with multiple layers of biomedical data, such as RNA, proteins, brain cells, and brain tissues \citep{zhao2016annual}.
   It is necessary to make greater efforts to collect and integrate multiple types of biomedical data and develop better statistical models for causal  analysis \citep{yu2022mapping}.  
   Clinical applications can also benefit from recent imaging genetic discoveries. For example, the combination of genetic polygenic risk scores and MRI could better predict the risk of brain diseases \citep{kauppi2018combining}. 
  
  \subsubsection{Causality  Research} \label{causal}

Causality research has received a lot of attention in neuroscience research \citep{friston2009causal,ramsey2010six,lindquist2012functional,yu2022mapping,sobel2020estimating,yu2022mapping,taschler2022causal,knutson2020implicating,zhao2021multimodal,le2019mapping,zhao2016annual,li2021bayesian}. Some important scientific questions in neuroscience include  how experimental stimuli affect brain function, how  different brain regions are causally linked for a specific task,  
how brain structure and function are causally linked with each other,    how brain structure mediates the relationship between 
  genetics and  clinical variables,    how brain mediates the relationship between 
  therapies/drugs and a clinical variable for brain-related diseases, and  what are the causal relationships between genetics, brain, health factors, and brain disorders.   Addressing these  questions raises serious challenges 
  in experimental design, data collection and integration, unobserved confounders,   SL methods for causal research, and causality validation, among others.   For instance, although 
 randomised controlled trials (RCTs) have been widely regarded as  the gold standard for causal discovery,  
 it might be inappropriate   to run RCTs in many neuroscience scenarios due to ethical or practical reasons. Therefore, 
one may have to draw causal conclusions  from existing observational data
under a series of `strict'  assumptions. 
 
 Causality research  can be roughly divided into causal discovery for determining causal relationships among a set of variables  and causal inference for estimating causal effects  deriving from a change of a certain variable over an outcome of interest in a large system \citep{imbens2015causal,pearl2009causality,greenland1999confounding,upadhyaya2021scalable,imbens2020potential}.  
  Causality research proceeds with  the development of the causal models 
  (e.g.,  the CGIC pathway in Figure \ref{fig:integration})
  for a set of  variables with possibly unobserved confounders.     
 Three main causal models include  the Bayesian network (BN) model based on a directed acyclic graph (DAG), the structural causal model (SCM) given a DAG, and the Rubin causal  model (RCM).    
 These causal models  complement  with each other and have their own pros and cons. 
 Under some conditions, SCM is a causal BN model, while  
RCM is logically equivalent to SCM \citep{pearl2009causality}.   
  The SCM and BN are more popular in computer science and epidemiology since they offer a graphical representation with reasonable interpretability and explainability.  In contrast,  RCM is very popular in statistics,  economics,  
and social sciences, since it is well connected with   experimental design and estimating causal effects.

 The causal discovery methods for causal BN (CBN) can be categorized into discrete space algorithms and continuous space algorithms \citep{upadhyaya2021scalable}.  
Traditional discrete space algorithms, including constraint-based and score-based
methods,  search for the optimal graph from a discrete space of candidate graphs by using either statistical tests or scores (e.g., Bayesian information criterion) 
to estimate the causal structure of DAG. In contrast, continuous space algorithms find an optimal graph from the continuous space of weighted
DAGs based on machine learning  algorithms.  Computationally, the complexity of traditional discrete space algorithms grows with  the number
of nodes in DAG, whereas continuous space algorithms are more scalable. 
Moreover,  causal discovery methods are designed to three types of data under analysis, including cross-sectional, time series, and longitudinal data. Distinguishing cross-sectional and time-series data is that  
there is a time component in time-series data so that events in the present cannot cause events in the past.  
The Granger causality method is one of the well-known methods for performing causal discovery for time-series data.

As an illustration, we consider  different causal discovery methods for using functional neuroimaging data (e.g., fMRI) to infer effective connectivity, which is 
a causal model of the interactions between different ROIs.  
  Different discrete space algorithms and their extensions have been used for effective connectivity 
   \citep{smith2011network}. 
Other statistical methods for effective connectivity include 
   Granger causality, dynamic causal models, structural equation models, state-space models, RCMs,   directed graphical models, and dynamic Bayesian network models, among others 
 \citep{friston2009causal,ramsey2010six,lindquist2012functional,sobel2020estimating}.  
 However,  most existing network methods suffer from large estimation errors for connection directionality \citep{li2021bayesian}.

We estimate the causal effect of a specific treatment $(X)$ over a certain outcome of interest $(Y)$ in two steps, including   
 (i): the study of
identification questions for $X\rightarrow Y$ and  (ii): estimation and inference methods for the causal effect $X\rightarrow Y$.  
 Specific  identification strategies for Step (i) include experimental design, 
adjustment/unconfoundedness, instrumental variables, difference-in-differences, regression discontinuity designs, synthetic control methods, and causal mediation analysis, among others.  
For instance,  it is common to use the frontdoor and backdoor criteria  to identify  valid adjustment sets 
\citep{pearl2009causality,upadhyaya2021scalable,imbens2020potential}.  
Causal inference algorithms only  work when  all common causes of  $X$ and $Y$ have been included in observational data, called {\it  causal sufficiency}, so  
controlling unobserved confounding requires a series of strong assumptions \cite{burgess2017review,zhu2020mendelian}.    
In Step (ii), SCM explicitly specifies all mediators, whereas RCM does not 
handle unspecified mediators in the outcome-generating model.

As an illustration, we consider the integration of multi-view data in ADNI to infer a hypothetical causal 
model for biomarker dynamics in AD pathogenesis presented in   
 \citep{jack2010hypothetical}. It starts from  AD risk genes to the abnormal deposition of $\beta$ amyloid (A$\beta$) fibrils, to increased levels of CSF tau protein, to  hippocampal atrophy, to declined cognitive symptoms and impairment, to AD.      Existing SL methods focus on  associations between different views,  but there is a growing interest in 
 delineating the {\it temporal causal relations} in Jack's causal model, say    \textit{the causal effect of hippocampal atrophy $(X)$ on behavioral deficits $(Y)$} \cite{yu2022mapping}. Our CGIC pathway is an approximation to Jack's causal model.  We need to check the causal sufficiency of $X$ and $Y$, which is most likely invalid in practice.  Although there are several popular identification strategies, including instrumental variables and the frontdoor criterion, for handling the issue of unobserved confounding, each of them has to make some strict assumptions. For instance, Mendelian randomization is an instrumental variable method, which selects a set of genetic variants $(G)$ as instruments to estimate the causal effect of $X\rightarrow Y$ \citep{sanderson2022mendelian}. It requires three key assumptions including   relevance,  independence, and  no horizontal pleiotropy. 
It can be implemented using  individual-level data in a single sample or summary data from two samples.  
Several popular instrumental variable estimation methods include the ratio
method, two-stage methods, likelihood-based methods, and semi-parametric methods \citep{burgess2017review,zhu2020mendelian}.
Furthermore, it is of great interest to build SCMs to link all variables in ADNI together and 
infer their time-varying causal relationships by extending causal mediation methods \citep{vanderweele2015explanation}.  This is motivated by delineating how most 
brain-related disorders  progress  and change adjusting for
temporal confounding by 
various health factors \citep{zhao2016annual}.

  \subsubsection{Predictive Models (PM)} \label{predictive}

 There is a large literature on the 
 development of SL methods for  building various predictive models in neuroscience research and clinical translation  \citep{kohoutova2020toward, davatzikos2019machine,hast:tibs:01,zhou2021review}.  
  The goal of predictive models is to use  a set of current and historical features in ${\bf x}$ to predict future  events in $Y$.  
Such development is motivated by identifying  biomarkers (e.g., neuroimaging) that could potentially aid detection, diagnosis, prognosis, prediction, and monitoring of disease status, among many others.   
As shown in Figure \ref{fig:integration}, the feature vector ${\bf x}$ in NDA may include neuroimaging, genetic, environmental and demongraphic variables and 
$Y$ is a low-dimensional vector consisting of cognitive scores,  diagnosis,  and survival time, among others.  
Despite how much progress has been established in academic settings recently, most predictive models have not been transferred to clinical practice in NDA.

A good predictive system  in NDA for clinical translation  includes 
  (i) a feature engineering pipeline to generate cost-effective and reliable biomarkers (e.g., blood) and perform high-quality data annotation,   
(ii) SL methods for training predictive models with high predictive capacity, robustness, and clarity for main NDA tasks, and (iii) a feedback loop to improve (i) and (ii).  
Developing a good predictive system requires 
appropriately handling (CT1)-(CT8), among which (CT4) needs more close attention.  
 Model (\ref{eq1}) emphasizes that  neuroimage data contain external  heterogeneity caused by  exogenous factors (e.g., 
 device, acquisition parameters) and internal  heterogeneity associated with downstream tasks for $Y$ \citep{liu2021statistical}. 
Specifically, "internal heterogeneity" refers to how diseased  regions may significantly vary across subjects and/or time in terms of their  number, size, degree, and location.   
 A good predictive system has to 
appropriately handle both external heterogeneity and internal heterogeneity in neuroimage data
through further developments in  (i) and (ii), among which (i) is the biggest bottleneck.

We discuss the pros and cons of existing  SL methods for predictive models in NDA.  
First, most existing supervised learning and variable selection methods \citep{hast:tibs:01}  are  sub-optimal 
  for predictive models in NDA due to the non-sparse effect of image biomarkers on $Y$ and (CT4) in neuroimaging data.
  Second, DL methods \citep{goodfellow2016deep} have achieved very promising results for handling pattern recognition problems, which include the issue of internal heterogeneity in neuroimaging data discussed above. Training good predictive models requires large-scale representative datasets with high-quality data annotation.   Third,  it is interesting to develop SL methods for causal predictive models, which use  causal thinking to improve prediction modelling, in NDA   \citep{imbens2015causal,pearl2009causality}. Specifically, we may test and validate the dynamic causal relationships in Figure \ref{fig:integration} based on observational data and then incorporate such causal findings to enable risk estimation under hypothetical interventions.

  \subsection{Challenges} 
  
  We have briefly reviewed the nine important PSA techniques above, but most of them are emerging fields and pose   many statistical challenges. 
First, the complexity of those large-scale neuroimaging-related data sets is too high for most research teams in both academia and industry. 
  It requires a close multidisciplinary  collaboration among  experts with strong skills in   statistics, biostatistics, epidemiology,  genetics/genomics,  engineering,  applied mathematics, machine learning, neuroscience, brain disorders,   imaging physics, and imaging analysis.   
Second, it is very difficulty  to appropriately process data across different domains with high quality, while controlling for potential bias 
  introduced during the preprocessing stage.  It requires the whole scientific community to work closely to test all major preprocessing tools by using well-designed synthetic and real datasets in terms of   reproducibility, generalizability, and reliability.  
 Third, it remains uncertain as to how to appropriately integrate data across different domains obtained from different studies and cohorts with possible  different study designs for unbiased data integration. Although one might attempt to integrate as many variables and studies as possible in a project, it would likely lead to serious biases in downstream data analyses and conclusions.  
 Fourth, it remains unclear how to appropriately and efficiently analyze neuroimaging related data sets with  multiple Vs (e.g., Volume, Velocity, Variety and Veracity), while ensuring algorithmic fairness. 
 Many existing statistical and machine learning models were
 developed before the era of big data, so they might make some strong assumptions that are inappropriate for neuroimaging related data sets 
 as discussed in Sections 2 and 4. We expect that there will be many novel SL methods   for NDA in the next decade.





\section{Supplementary Material}
A supplementary file document was included.

\section{ACKNOWLEDGMENTS}
This research was partially supported by U.S. NIH grants MH086633 and MH116527. We acknowledge Prof. Tianming Liu and Miss 
Wyliena Guan for many insightful comments. 

\section{DISCLOSURE STATEMENT}
The authors are not aware of any affiliations, memberships, funding, or financial holdings that might be perceived as affecting the objectivity of this review.

\bibliographystyle{chicago}
\bibliography{imageref,refsg,causal}

\pagebreak
\begin{center}
  \title{\Large\bf Supplementary Material for ``Statistical Learning Methods for Neuroimaging Data Analysis with Applications''}
  \end{center}
    \maketitle
 \begin{center}
   \author{\large Hongtu Zhu$^{1}$,  Tengfei Li$^2$, and Bingxin Zhao$^3$ \\
   \vspace{10pt}
   $^1$Departments of Biostatistics, Statistics, Genetics, and Computer Science and Biomedical Research Imaging Center,  University of North Carolina, Chapel Hill \\
   $^2$Departments of Radiology and Biomedical Research Imaging Center,  University of North Carolina, Chapel Hill \\
   $^3$Department of Statistics and Data Science, University of Pennsylvania
   }\\
   \end{center}

\spacingset{1.7} 

The supplementary file include: 
\begin{itemize}
    \item Supplementary table S1.
\end{itemize}
\begin{table}[!h]
\begin{center}\caption{Summary of key information for eight neuroimaging modalities}
\footnotesize
\begin{tabular}{ |p{1.2cm}||p{2.8cm}|p{2cm}|p{2.5cm}|p{3.3cm}|p{2.5cm}|  }
 \hline
Modality&Tracer& Resolution&Feature &Use&Software\\
 \hline
 sMRI (T1, T2)   &Fluid characteristics of different tissues& 0.5-1 mm   &Cortical thickness, cortical  folding,
 sulcal depth, voxel-based morphometry, 
regional volumes and shape& Measure brain cortical/subcortical structural changes for diagnosis/staging/follow-up of disease/brain development. &   Freesurfer, ANTs, FSL, SPM, AFNI, Hammer, BRAINVisa, BrainSuite\\ \hline
 DWI &Brownian motion of water molecules within voxels &   1.25-3 mm  &Fractional anisotropy, axial/radial/mean diffusivity, DKI/ NODDI parameters, structural connectivity & Delineate tumors, suspected acute ischemic brain injury,  intracranial infections, masses, trauma, and edema; map structural connectome in research.   &FSL, ~~~~~~Mrtrix, AFNI, TrackVis, Camino, TORTOISE, slicerDMRI, Dipy, CAMINO, DSIStudio\\ \hline
 fMRI & Blood-oxygen-level-dependent (BOLD)  response in blood flow  associated with brain function&3-4 mm (spatial); 1-3 s (temporal) &Beta image, functional connectivity, weighted and binary network metrics& Brain activity mapping under tasks, 
 brain abnormalities detection, pre-operative  brain functional mapping. & SPM, FSL, AFNI, CPAC, FuNP\\ \hline
 PET    & Emissions from  radioactive tracers&4-5 mm &Standard uptake ratio & Reveal metabolic/ biochemical functions of tissues/organs and abnormalities in brain neurophysiology/ neurochemistry& NiftyPET, SPM, Metavol, NEUROSTAT, APPIAN, kinfitr, LIFEx, Pypes, SPAMALIZE\\ \hline
 CT& X-ray attenuations by different tissues inside the body&   Tens of nanometres-5 mm & Local and regional volumetric/thickness measures, tumor features  &Diagnosing a range of conditions: abnormal blood vessels, brain atrophy, hemorrhage, swelling, stroke, tumors &ITK, SPM, PACS, Velocity, scenium, LIFEx\\ \hline
 EEG& Electrical field  produced by neuron electrical activity& 7-10 mm  &Event-related potentials, connectivity/network measures, spectral content& Diagnosis and treatment of 
brain tumors, damage, dysfunction and disorders &EEGLAB, MNE, ELAN, FieldTrip, NUTMEG, BrainVoyager, SPM\\ \hline
 MEG&Magnetic field  produced by neuron electrical activity, including tangential components of  postsynaptic intracellular currents& 2-3 mm  &Similar derived measures with EEG& Identification of brain functional areas (centers of sensory, motor, language and memory activities), precise location mapping 
  of the source of epileptic seizures &EEGLAB, MNE, ELAN, FieldTrip, NUTMEG, BrainVoyager, SPM\\ \hline
 fNIRS&Changes in cortical BOLD response associated with brain function& 650-900 nm (spatial); milliseconds (temporal) &Similar derived measures with EEG and fMRI& Study normal and pathological brain physiology in infants/children&Homer2, Homer3, FNIRSOFT, OPENFNIRS, ICNNA, nirsLAB \\ \hline
 \hline
\end{tabular}  
\end{center}\label{Table:modality}
\end{table}

\end{document}